\documentclass[12pt%
              ,a4paper%
              ]{article}

\usepackage{SSSTSiQFT-gCFT_aux}

\usepackage{Include}
\usepackage{color}

\newcommand{\be}{\begin{equation}}
\newcommand{\ee}{\end{equation}}
\newcommand{\ba}{\begin{eqnarray}}
\newcommand{\ea}{\end{eqnarray}}

\newcommand{\bd}{\begin{description}}
\newcommand{\ed}{\end{description}}

\renewcommand{\iota}{{\bf 1}}

\def\rellow#1#2{Mathrel{Mathop{\kern 0pt #1}\limits_{#2}}}

\begin{document}

\title{
\vskip-3cm{\baselineskip14pt
\centerline{\normalsize \hfill DESY 04-138}
\centerline{\normalsize \hfill hep-th/0410277}
\centerline{\normalsize \hfill \strut}
}
\vskip1.5cm
Simple Space-Time Symmetries:\\ 
Generalizing Conformal Field Theory}
\author{Gerhard Mack and Mathias de Riese\footnote{Funded by the German National Academic Foundation}\\
   II. Institut f\"ur Theoretische Physik, Universit\"at Hamburg
\date{January 25,  2006}}

\maketitle

\textbf{Abstract} We study simple space-time symmetry groups $G$ which
act on a space-time manifold $\Spacetime = \quotient{\rLG}{\Homstab}$ which admits a
$G$-invariant global causal structure.
  We classify pairs $(G,\Spacetime)$ which
share the following additional properties of conformal field theory:
1) The stability subgroup $\Homstab$ of $\Hid\in \Spacetime $ is
the identity component of a parabolic subgroup of $G$, implying
factorization $\Homstab=\Lorentz\Dil\Spc$, where $\Lorentz$ generalizes
Lorentz transformations, $\Dil$ dilatations, and $\Spc$ special
conformal transformations.  2) special conformal transformations
$\onespc\in\Spc$ act trivially on tangent vectors $v\in
\Tang{\Hid}{\Spacetime}$. The allowed simple Lie groups $\rLG$ are the
universal coverings of $\LG{SU(m,m)},\LG{SO(2,\SOdim)},\LG{Sp(l,\R)},
\LG{SO*(4n)}$ and $\LG{E_7(-25)}$ and $\Homstab$ are particular
maximal parabolic subgroups.
They coincide with the groups of fractional linear transformations of Euklidean Jordan algebras whose use as generalizations of Minkowski space time was advocated by G\"unaydin.
 All these groups $\rLG$ admit positive
energy representations.  It will also be shown that the classical
conformal groups $\LG{SO(2,\SOdim)}$ are the only allowed groups which
possess a time reflection automorphism $\timerev$; in all other cases
space-time has an intrinsic chiral structure.

\section{Introduction} 
\label{sec:intro}
Although the Poincar\'e group is not semi-simple, it is of interest to
study also quantum field theories (QFT) with semi-simple space-time
symmetry groups $\rLG$.  It is necessary for the existence of QFT
satisfying conventional principles
that $\rLG$ acts on a space-time manifold $\Spacetime$ which admits a
$\rLG$-invariant global causal structure, implying a notion of spacelike,
positive or negative timelike tangent vectors, and $\Spacetime$ should
be globally hyperbolic.  If the existence of a ground state is
demanded, $\rLG$ should admit positive energy representations.

We are only interested in cases where $\rLG$ acts transitively on
$\Spacetime$, hence
\begin{equation}
\elabel{manifoldquotient}
  \Spacetime = \quotient{\rLG}{\Homstab}\;,
\end{equation}
where $\Homstab $ is the stability
subgroup of some point $\Hid \in \Spacetime$.

Conformal field theories in arbitrary dimension $\SOdim$ are the
classical examples.  It had been believed for a long time that
conformal symmetry had causality problems because special conformal
transformations in Minkowski space could interchange spacelike and
timelike, but the issue was clarified by L\"uscher and Mack
\cite{LueMack:COMMP-41-203}. Causality (in the sense of global hyperbolicity) 
is satisfied when one considers
the action of the universal covering $\rLG$ of the conformal group
$\LG{SO(2,\SOdim)}$ on an $\infty$-sheeted covering $\Spacetime$ of
compactified Minkowski space.

In the conformal case the pair $(\rLG, \Spacetime)$ has the following
two additional properties, which can be formulated for general
semi-simple Lie groups.

\begin{assumption}[ass:1]
  The stability subgroup $\Homstab$ of $\Hid\in \Spacetime $ is the
  identity component of a parabolic subgroup of $\rLG$, implying
  factorization $\Homstab=\Lorentz\Dil\Spc$, where elements of $\Lorentz$
  generalize Lorentz transformations, $\Dil$ dilatations, and $\Spc$
  special conformal transformations.
\end{assumption}

\begin{assumption}[ass:2]
  Special conformal transformations $\onespc\in\Spc$ act trivially on
  tangent vectors to $\Spacetime $ at $\Hid$.
\end{assumption}

It is essential that one divides by the identity component of the
parabolic subgroup and not by the whole group. Otherwise causality
will be destroyed.  By their definition, parabolic subgroups contain
the whole center of $\rLG$ which is infinite in our case, implying
that the parabolic subgroup has infinitely many connected components.
If one were to divide by the center, space-time $\Spacetime$ would
become compact -- compactified Minkowski space in the conformal case
- and therefore not globally hyperbolic.

We wish to classify the pairs $(\rLG,\Spacetime)$ such that
 $\Spacetime $ admits a global
causal structure, and the two properties \rst{ass:1} and
\rst{ass:2} just mentioned are true, with appropriate groups $\Lorentz$,
$\Dil$, $\Spc$.  The definition of a parabolic subgroup implies that
$\Dil$ is non-compact abelian, the Lie algebra of $\Spc$ is nilpotent,
and $\Lorentz$ is a reductive Lie algebra whose elements commute with
those of $\Dil$.

The results which are to be proven in this paper are stated in 
 \rsec{summary}. The classification is in \rst{listOfCausals} and 
\rfig{ParblStruct}.

  It turns out that under the stated assumptions global hyperbolicity of
  $\Spacetime$ implies that $\rLG$ possesses positive energy
  representations, see \rst{theo:4} below and the subsection following it.

We also examine the question which of the groups $G$ admit a time reflection
 automorphism.

\section{Summary of Results}
\label{sec:summary}
\subsection{First Implications of Assumptions 1 and 2}

The assumptions 1 and 2 go a long way towards satisfying
causality requirements, if $\rLG$ is connected and simply connected.
For groups which are not simply connected, problems like closed
timelike paths may appear.  They are eliminated by passing to the
universal covering $\rLG$ of the group.

To state the results precisely, we must recall some structure
theory. 

The Lie algebra $\rLA$ of a non-compact Lie group $\rLG$ admits a Cartan decomposition,
\begin{equation}
\elabel{Cartan} 
  \rLA=\Ck + \Cp  \ .
\end{equation}
The center $\Ctr $ of $\rLG$ is contained in the subgroup $\aCk$ of $\rLG$ with Lie algebra $\Ck $ and $\quotient{\aCk}{\Ctr} $ is the maximal compact subgroup of $\quotient{\rLG}{\Ctr}$. We will call $\Ck$ the maximal compact subalgebra of $\rLA$ for short. But note that $\aCk$ is not compact if the center $\Ctr $ is infinite. 

 If $\Homstab = \Lorentz\Dil\Spc$ is the identity component of a
parabolic subgroup of $\rLG$, there exists an inner involutive
automorphism $\autspctrl$ of $\rLG$ which carries $\Lorentz$ to $\Lorentz$, $\Dil$
to $\Dil$ and maps $\Spc$ to an isomorphic subgroup $\Trl$ such that the
Lie algebra $\gLA$ admits a direct sum decomposition
\begin{equation}
\elabel{Bruhat} 
  \rLA=\trl\LAsum (\dil \LAsum \lorentz ) \LAsum  \spc
\end{equation}
where $\trl$ is the Lie algebra of $\Trl$ etc. (generalized Bruhat
decomposition). $\dil$ is abelian, $\lorentz$ is reductive, $\trl$ and
$\spc$ are nilpotent, $\dil \subset \Cp$,  and the elements of $\dil$ and of $\lorentz$
commute. Furthermore,
\begin{equation} 
\elabel{actionOnN+}
  \comm{\dil}{\trl}\subseteq \trl\;, \quad \comm{\lorentz}{\trl}\subseteq \trl \;,
\end{equation} 
and similarly for $\spc$.
 
\begin{theorem}[theo:1]
\label{theo:1}
  It follows from the assumptions 1 and 2 that the corresponding
  generalized Bruhat decomposition \req{Bruhat} of $\rLA$ satisfies
\begin{equation}
\elabel{assumption:1}
  \comm{\trl}{\spc} \subseteq \dil \LAsum \lorentz\;.
\end{equation}
Conversely, suppose that $\Homstab$ is the identity component of a parabolic
subgroup. Then \req{assumption:1} implies that $\spc$
acts trivially on tangent space $\Tang{\Hid}{\Spacetime}$.
\end{theorem} 

In fact, more is true:


\begin{theorem}[abelN1dimA]
  Assuming $\rLA$ is simple, the commutation relations
  \req{assumption:1} imply that $\trl$ and $\spc$ are
  commutative and $\dil$ is 1-dimensional. 
\end{theorem}

In this sense the groups $\rLG$ are very similar to the conformal
group, apart from the generalized Lorentz group. In particular, a
generalized Poincar\'e group $\Lorentz\Trl$ exists which is the
semi-direct product of an abelian group $\Trl$ and a generalized
Lorentz group $\Lorentz$. 

\subsection{Causal Structure}

Now we are ready to formulate the first main result:


%
\begin{theorem}[theo:4]
  Suppose that $\rLG$ is a non-compact simple real Lie group with Lie
  algebra $\gLA$ and assumptions 1 and 2 are satisfied. Then
  \begin{itemize}
  \item[i)] Assuming 
the complexification $\cpl{\rLA}$ of $\rLA$ is simple and 
$\Lorentz$ is semi-simple or trivial, $\Spacetime$ admits an
    $\rLG$-invariant infinitesimal causal structure if
    and only if the tangent space $\Tang{\Hid}{\Spacetime}$ to
    $\Spacetime$ at $\Hid$ possesses a non-zero vector which is
    invariant under the rotation group $\Lorentz \cap \aCk$.
  \item[ii)] $\Spacetime$ is globally hyperbolic if and only if the maximal
    compact Lie subalgebra $\Ck $ of $\gLA$ has nontrivial center
    $\Ckc$, $\Ckc$ is not contained in the Lie algebra $\lorentz$ of
    $\Lorentz$, 
 and the corresponding Lie group $\Ctr_\Ck \simeq \R$.
    \com{Die Existenz des Zentrums braucht nur Assumption 1}
  \end{itemize}
\end{theorem}
By definition, global hyperbolicity implies the existence of a 
$\rLG$-invariant infinitesimal causal structure
 (i.e. of light cones with appropriate properties) see section 
\ref{sec:globalcausality}.  It will follow from the final classification 
that the assumption  in the if part of ii) that  $\Ckc$ is not contained in 
$\lorentz$  is automaticaly fulfilled.

Let us explicitly state that the universal covering group $\rLG$ of
$\LG{SO(2,1)}=\quotient{\LG{SL(2,\R)}}{\Z_2}=\quotient{\LG{SU(1,1)}}{\Z_2}$
is covered by the theorem. In this case $\lorentz$ is trivial, and
$\Spacetime=\R$. The invariant cones defining the infinitesimal causal
structure are half lines $\R_\pm$.

Let us agree for the sequel of this paper to count trivial $\lorentz$ as 
semi-simple. 

Let us also note that $\rLA $ simple implies that $\cpl{\rLA}$
is simple, unless $\rLA$ is a complex Lie algebra, and in the latter case,
$\Ck $ does not have nontrivial center.
\subsection{Positive Energy Representations}
\label{sec:IntroPERep}

Positive energy representation means that there exists a generator $\I
\energy$ in the Lie algebra $\rLA$ of $\rLG$ such that $\energy$ has
positive spectrum ($\energy\geq 0$) in the representation. The
non-compact simple Lie groups $\rLG$ possessing positive energy
representations have been classified long ago \cite{Luscher:DESY-75-51}.
Let $\Cent$ be the center of $\rLG$ and $\quotient{\aCk}{\Cent}$ the
maximal compact subgroup of $\quotient{\rLG}{\Cent}$. 
  $\rLG$ possesses positive energy representations if
$\quotient{\aCk}{\Cent}$ has a $\LG{U(1)}$-factor, i.e.  $\Ck $ has
nontrivial center $\Ckc$.  If $\rLA$ is simple, $\Ckc$ is
1-dimensional and $\energy \geq 0$ in positive energy representations, for
a generator $iH_0\in \Ckc$.

By the second part of \rst{theo:4}, global hyperbolicity
of $\Spacetime$ implies that this criterion for the existence of
positive energy representations is satisfied.

It follows that the center $\Ctr$ of the simply connected group $\rLG$
is infinite, containing a factor $\Z$. The simple groups $\rLG$ with
positive energy representations are coverings of
\begin{equation}
 \elabel{groupsWithPosEnergyReps} 
 \LG{SU(m,n)},\ \LG{SO(2,\SOdim)},\ \LG{Sp(l,\R)},\ \LG{SO*(2l)}, \ \LG{E_6(-14)},\ \LG{E_7(-25) }
\end{equation}

 In the conformal case, the rotation invariant
generator $P_0$ of Poincar\'e-transla\-tions is also positive, because
it is limit of elements conjugate to $\energy$ under dilatations, and
conversely, $\energy=P_0+\autspctrl(P_0)$. This result generalizes. In
particular, $\I\energy\in(\trl\VSsum\spc)\cap\Ck$, and it lies in the
commutant of $\lorentz\cap \Ck$.


\subsection{Classification}
A classification is obtained from \rst{theo:1} by inspection of root
systems as will be described below. The result is as follows.

\begin{theorem}[listOfCausals]
  The pairs $(\rLG,\Spacetime)$ such that $\rLG$ is a simple connected
  simply connected Lie group
  and $\Spacetime=\quotient{\rLG}{\Homstab}$ where
  $\Homstab=\Lorentz\Dil\Spc$ is identity component of a parabolic
  subgroup of $\rLG$ satisfying assumption 1 and 2 and 
    such that $\Spacetime$ carries a $\rLG$-invariant global causal structure
  are as follows.
  
  $\rLG$ is the universal covering of one of the groups
  $\LG{SU(m,m)}$, $\LG{SO(2,\SOdim)}$, $\LG{Sp(l,\R)}$,
  $\LG{SO*(4n)}$, and $\LG{E_7(-25)}$, with $m\ge 1, \SOdim \ge 3,
  l\ge 2, n\ge 2 $, and $\Homstab$ is a maximal parabolic subgroup.
  $\Homstab$ are uniquely determined by $\rLG$ up to conjugation. In all
  these groups, $\I\energy$ is never a generator of $\Lorentz$.
\end{theorem}

The following \rfig{ParblStruct} lists the Lie algebras $\rLA$,
together with the Lorentz subalgebra $\lorentz$ and the semi-simple
part $\Cks$ of the Lie algebra $\Ck=\Cks\LAsum\Ckc$ of the maximal
compact subgroup $K/\Ctr$ of $G/\Ctr$. $r$ is the split rank of $\rLA$, i.e. the maximal number of non-compact generators in a Cartan subalgebra.
 The remaining entries determine
the parabolic subgroup and $H_0$, as follows.

It is recalled in \rsec{parabolic} how the parabolic subgroups of
$\rLG$ are classified by proper subsets $\Iroots$ of the set of simple
restricted roots. For maximal parabolic subgroups, $\Iroots$ lacks a
single simple restricted root $\srrext$, which is restriction of a
root $\srext$. The entry $\srrext$ of the table gives the
corresponding node in the Dynkin diagram of the restricted root system
$\rRSys$ of $\rLA$, \rfig{SatakeDiag}.  Nodes are enumerated $1,2,...$
from top to bottom and from left to right.
The last column of the table contains information about $\energy$ and will be explained at the beginning of \rsec{HandTime}.

\begin{mytable}{6}{&&&&&}{|c|c|c|c|c|c|c|} %
  {Type        & $r$ & $\srrext$ & $\lorentz$    & $\dim\Spacetime$ & $\Cks$ & Node  }
  $\LA{SU(m,m)}$    & $m$ & $m$       & $\LA{Sl(m,\C)}$  & $m^2$ & $\LA{SU(m)}\oplus \LA{SU(m)}$ & $m$
  \END
  $\LA{SO(2,\SOdim)}$ & $2$ & $2$     & $\LA{SO(1,\SOdim -1)}$ & $\SOdim$ & $\LA{SO(D)}$ & $2$
  \END
  $\LA{Sp(l,\R)}$   & $l$ & $1$       & $\LA{Sl(l,\R)}$   & $\frac 1 2 l(l+1)$ & $\LA{SU(l)}$
& $1$
  \END
  $\LA{SO^*(4n)}$   & $n$ & $1$       & $\LA{SU^*(2n)}$   & $2n(n-1)$ & $\LA{SU(2n)}$
& $1$
  \END
  $\LA{E_{7(-25)}}$ & $3$ & $3$       & $\LA{E_{6(-26)}}$ & $27$ & $\LA{E_6}$
 & $7$
  \END 
  \hline \mycaption{ParblStruct}{The Lie algebras with positive energy
    representations which admit parabolic subalgebras satisfying
    assumptions 1 and 2. $m\ge 1$, $\SOdim\ge 3$, $l\ge 2$, $n\ge 2$.
    Their split rank $r$, the Lorentz subalgebra $\lorentz$ and
    semi-simple part $\Cks$ of the maximal compact subalgebra $\Ck$ of
    $\rLA$, and the dimension of $\Spacetime$ are also listed. The
    remaining entries determine the parabolic subalgebra and
    $\energy$, see text.}
\end{mytable}

\subsection{Interesting Observations}
\label{sec:obs}
It is an important observation that $\lorentz$ and $\Cks$ have
isomorphic complexification, and therefore also $\lorentz\LAsum\dil$
and $ \Ck = \Cks \oplus \Ckc$.  Their finite dimensional
representations can therefore be identified, by Weyls unitary trick.
This permits to relate lowest weight representations of $\rLA$, which
are determined by the lowest weight of a \begin{corr}finite dimensional\end{corr}
representation of $\Ck$, with induced representations of $\rLG$ on
$\Spacetime$, which are determined by finite dimensional
representations of $\lorentz\LAsum\dil$, see the Outlook. 
This fact and its implications were noted before by G\"unaydin
\cite{GAdSCFT1,GAdSCFT2}.
 Lowest weight representations of $\rLA$ are positive energy
representations.  Such representations have been studied long ago by
Harish-Chandra \cite{HC1,HC2}.

There are similarities among the Lorentz groups. With the remarkable
exception of conformal theories in odd dimensions $\SOdim$, nonzero
$\lorentz$ are real 
\begin{del}forms \end{del}%
\begin{corr}Lie algebras \end{corr}%
which owe their existence to a symmetry of their Dynkin diagrams. They
possess no compact Cartan subalgebra, as can be seen from the fact
that their maximal compact subalgebras $\rota=\lorentz\cap\Ck$ have
lower rank. For instance, $\rLA=\LA{E_7(-25)}$ has $\rota=\LA{F_4}$.
The Lie algebra $\LA{SU(1,1)}$ is a degenerate case with $\lorentz=0$.

Among the listed groups $\rLG$ there is one non-compact real form
$\LG{E_7(-25)}$ of an exceptional group. Space-time $\Spacetime$ has
dimension $27$ in this case. From the point of view of string theory
this is one dimension too much for a bosonic theory.  But then, the
$11$ dimensions of $11$-dimensional supergravity are also one too much
for a supersymmetric theory.
 
The exceptional group $\LG{E_6(-14)}$ also possesses positive energy
representations, but it has no parabolic subgroup which fulfills our
requirements.

Our assumptions amount to requiring that $\Spacetime$ is a
generalization of (an $\infty$-sheeted covering of compactified)
Minkowski space.  In the outlook we will mention some further quantum
field theoretic motivation for this assumption.

In a series of papers starting in 1975, G\"unaydin proposed the idea of using
Jordan algebras, and more generally Jordan triple systems, to define 
generalized space times and corresponding symmetry groups 
\cite{G1975,GConf1,GConf2,G1993,GAdSCFT1,GAdSCFT2}.
The list of groups which satisfy our selection criteria, which include
 global causality, coincides with the groups of conformal transformations (fractional linear transformations)  of Euklidean Jordan algebras  
\cite{G1993,Bertram:GJLS-00}. They generalize Minkowski space. This means that the Minkowski subspace of $\Spacetime$ can be made into a commutative
 nonassociative algebra. It does not admit a global causal structure nor is it
 a homogeneous space for $G$. The universal cover $\Spacetime$ of its 
compactification has both properties but is not an algebra.

All the simple groups with positive energy representations, including $E_{6(-14)}$ and $SU(m,n)$ with $m\neq n$ are conformal groups of Hermitean Jordan triple systems \cite{G1993}.

\subsection{Time Reflection}

We will derive one further result: It is a prerequisite for time
reflection symmetry of a $\rLG$-invariant theory that time reflections
$\timerev$ can act on $\Spacetime$, and this induces a time reflection
automorphism of $\rLG$.

Global hyperbolicity of $\Spacetime$ implies that it is homeomorphic to 
 $\R\times\Cauchy$, and interpretation of $T$ as a time reflection is
 understood to require that it reflects $\R$ and acts trivially on
 $\Cauchy$ for appropriate choice of $\Cauchy $ (``space''). .

\begin{theorem}[theo:timereversal]
  Of all the groups listed in \rst{listOfCausals}, only the conformal
  groups $\LG{SO(2,\SOdim)}$ admit a time reversal automorphism.
  
  The covering of the group $\LG{SU(1,2)}$ is the only other group
  with positive energy representations which admits a time reversal
  automorphism.
\end{theorem}
A $PT$-automorphism always exists. Therefore this means that in the
other cases, space-time has an intrinsic chiral structure.

\section{Action of $\Homstab$ on the tangent space  $\Tang{\Hid}{\Spacetime}$}
\begin{del}
  \bf \"Uberschrift war: Action of the stability group $\Homstab$ on
  tangent space $\Tang{\Hid}{\Spacetime}$
\end{del}

Denote the differentiable action of $g\in G$ on
$\Spacetime=\quotient{\rLG}{\Homstab}$ by $\Homact(g)$,
\begin{equation} 
\elabel{rhoact}
  \Homact(g)\coset[\Homstab]{\xInG} = \coset[\Homstab]{g\xInG} \quad\text{ for }\xInG\in\rLG\;.
\end{equation}
\com{$\coset[\Homstab]{\xInG}$ k\"onnen wir auch gerne $\xInG\Homstab$
  schreiben. Daf\"ur habe ich den Befehl \texttt{$\backslash$coset},
  den ich schnell \"andern kann. Diese Notation ist sehr verbreitet
  und wird auch in den Mathe f\"ur Physiker Vorlesungen verwendet.}

Since $\Homact$ determines a map of smooth curves, it induces a map
$\homact$ of tangent spaces, known as the derivative of $\rho$. The
action at $x\in \Spacetime $ will be denoted~by~$\homact^x$,
\begin{equation} 
\elabel{derivative} 
  \homact^x(g): \Tang{x}{\Spacetime} \mapsto \Tang{\Homact(x)}{\Spacetime} 
\end{equation}  

Given that $\Spacetime$ is the homogeneous space
$\quotient{\rLG}{\Homstab}$, the tangent space
$\Tang{\Hid}{\Spacetime}$ at $\Hid=\coset[H]{\Gid} $ can be identified with
the quotient vector space of the Lie algebras $\gLA $ and $\homstab$
of $\rLG$ and $H$,
\begin{equation} 
\elabel{tangSpace}
  \Tang{\Hid}{\Spacetime} = \quotient{\gLA}{\homstab} \ . 
\end{equation} 
The action \req{rhoact} of $g$ on $\Spacetime$ restricts to an action
of $h\in \Homstab$ on $\Spacetime$, $\Homact(h)\coset[\Homstab]{\xInG}
= \coset[\Homstab]{h\xInG h^{-1}}$.  It follows from this and from the
identification \req{tangSpace} that the induced action of
\begin{del}the Lie algebra $\homstab$ of \end{del}%
$\Homstab$ on $ \Tang{\Hid}{\Spacetime}$ is given by
\begin{equation}
\elabel{tangAct}
  \homact^\Hid(h)(\coset[\homstab]{X})=\coset[\homstab]{\Ad_{\rLG}(h)X}\;\text{, where } h\in\Homstab,\;X\in\rLA \;, 
\end{equation}
and $\Ad_{\rLG}$ is the adjoint representation of $\rLG$ acting on
$\rLA$.

Since $\homstab$ is an invariant subspace of $\rLA$ under the
restriction of the adjoint representation $\Ad_{\rLG}$ of $\rLG$ to
$\Homstab$, we see that
$\Tang{\Hid}{\Spacetime}=\quotient{\rLA}{\homstab}$ carries the
quotient representation.

This result can be phrased in another way which is a generalization of
a result used in the theory of causal symmetric spaces
\cite{FarautOlafsson:CSSSGHA-95}. 
\begin{del}
  Since our groups are connected Lie groups, we may pass between the
  action of group and Lie algebra on finite dimensional vector spaces
  without special mention.
\end{del}

\begin{lemma}[repONq]
  Choose any subspace $\homrest$ of $\rLA$ such that $\rLA$ is a
  direct sum, $\rLA=\homstab\VSsum\homrest$.  Let $\prj{\homrest}$ be
  the corresponding restriction to $\homrest$, viz.
  $\prj{\homrest}(Y+Q)=Q$ for \mbox{$Q\in \homrest, Y\in \homstab$.}
  The map
  \begin{equation}
    \elabel{intertw}
    \homact_\homrest:h\mapsto\prj{\homrest}\circ\Ad_{\rLG}(h)
  \end{equation}
  defines a representation of $\Homstab$ on $\homrest$ which is
  equivalent to its representation $\homact^\Hid$ on the tangent space
  $\Tang{\Hid}{\Spacetime}$
\end{lemma}
In favorable cases, $\Ad_{\rLG}(h), h\in \Homstab$ will map $\homrest$
into itself.  In this case the projector can be omitted.  We shall
also have occasion to consider cases where the projector cannot be
omitted.

\begin{del}
  We shall study the structure of parabolic subgroups $\Homstab$ of
  $\rLG$ in more detail later on. For now let us assume that its Lie
  algebra $\homstab$ has the following properties which hold in this
  case.
  \begin{equation} 
    \elabel{}
    \homstab =(\ICSA \LAsum \Ictz ) \LAsum  \Inilp[-] 
  \end{equation}
  where $\ICSA$, $\Ictz$ and $\Inilp[-]$ are subalgebras obeying
  \begin{equation} 
    \elabel{actionOnN-}
    [\ICSA, \Inilp[-]]\subseteq \Inilp[-], \quad 
    [\Ictz, \Inilp[-]]\subseteq \Inilp[-]
  \end{equation}
  and there is an involutive automorphism $\theta$ of $\gLA$ which
  carries $\ICSA$ to $\ICSA$, $\Ictz$ to $\Ictz$ and $\Inilp[-]$ to
  $\Inilp[+]$ such that $\gLA$ admits the generalized Bruhat
  decomposition \req{Bruhat}.  Eqs.\req{actionOnN+} follow from
  \req{actionOnN-} by applying $\theta$.
\end{del}
\com{War doch schon genau so in der Introduction}

\begin{proof}[ of \rst{theo:1}]\\
  According to the generalized Bruhat decomposition \req{Bruhat},
  $\rLA= \trl \LAsum \homstab $, i.e. we may choose $\homrest=\trl$ in
  \rst{repONq}. It follows from \req{tangAct} that trivial action of
  $\spc$ on $\Tang{\Hid}{\Spacetime} \simeq \homrest$ is
  equivalent to $\comm{\spc}{\trl}\subseteq\homstab$. Applying
  the involutive automorphism $\autspctrl$ to both sides it follows that
  $\comm{\trl}{\spc}\subseteq\autspctrl(\homstab)$.  Since $\homstab
  \cap \autspctrl(\homstab)= \dil\LAsum\lorentz$, \rst{theo:1}
  follows.
\end{proof}

Choosing $\homrest=\trl$ in \rst{repONq}, it simplifies because of
properties \req{actionOnN+}.
\begin{lemma}[AssumpSimple]
  Assuming trivial action of $\Spc$ on tangent space
  $\Tang{\Hid}{\Spacetime}=\quotient{\rLA}{\homstab}\simeq\homrest$,
  the assertion of \rst{repONq} simplifies for the choice
  $\homrest=\trl$ to
  \begin{equation}
    \elabel{}
    \homact_{\trl}(nl)= \Ad_{\rLG}(l)
  \end{equation}
  for $n\in\Spc, l\in\Lorentz\Dil$.
\end{lemma}
%

\section{Parabolic subalgebras and subgroups}
\label{sec:parabolic}

Parabolic subgroups $\IParbl[]=\ICtzI[]\aICSApE[]\aInilpE[]{-}$ of $\rLG$ 
play a
central role in the representation theory and harmonic analysis on
simple non-compact Lie groups $\rLG$ \cite{Warner:HASSLGI-72}. They are
used to construct induced representations of $\rLG$ on
$\quotient{\rLG}{\IParbl[]}$, of which $\Spacetime$ is a covering space. 
These representations are  induced by
finite dimensional representations of $\IParbl[]$
which are trivial on $\aInilpE[]{-}$.  It is known that all
irreducible representations of $\rLG$ are subrepresentations of such
induced representations. The unitary positive energy representations
of the conformal group in 4 dimensions were constructed in this way in
\cite{Mack:COMMP-55-1}. In cases of interest here, neither $\ICtzI[]$ nor 
$\IParbl[]$ are connected. We will need their identity components 
$M$ and $\Homstab= M\aICSApE[]\aInilpE[]{-}$, and their Lie algebras. 

Nonisomorphic parabolic subgroups will be labelled by $\srsubset $ as 
explained below. Our assumptions will require a special choice of $\srsubset$, and the parabolic subgroup mentioned in \rsec{summary} correspond to this 
choice.

\subsection{Minimal Parabolic Subalgebras}

Remember the Cartan decomposition $\rLA=\Ck + \Cp$ of the Lie algebra of 
$\rLG$. 
Let $\ICSAp$ be a subspace of $\Cp$ which is maximal subject to the
condition that its elements commute. $\ICSAp$ may be extended to a
Cartan subalgebra $\ICSAk\oplus\ICSAp$ of $\rLA$ with $\ICSAk\subset\Ck$.
This is called a maximally non-compact Cartan subalgebra. Consider
roots $\rta\in\RSys$ for this Cartan subalgebra. They are linear maps
from the Cartan subalgebra $\ICSAk\oplus\ICSAp$ to $\C$ and as such
may be restricted to $\ICSAp$. %
\begin{del}Denote the restriction map by $r$. \end{del}%
The resulting linear maps $\rtl$ on $\ICSAp$ are called restricted
roots and form the restricted root system $\rRSys$.

\com{Im folgenden habe ich versucht zu reparieren, dass kein
  Unterschied zwischen reellen und komplexen root spaces gemacht
  wurde. Ist das verstaendlich?}

The root spaces $\RSp{\rta}$ for roots $\rta\in\RSys$ are
one-dimensional subspaces of the complexification $\cpl{\rLA}$ of
$\rLA$ and obey
\begin{equation} 
\elabel{plusRoot}
  \comm{\RSp{\rta}}{\RSp{\rtb}}= \RSp{\rta+\rtb}
\end{equation}
as an equality (\cite{Wallach:HAHS-73}, Lemma 3.5.10), with
$\RSp{\rta+\rtb}=0$ if $\rta+\rtb$ is not a root.
 
The root spaces may be composed to restricted root spaces
$\cpl{\RSp{\rtl}}=\sum\RSp{\rta}$, where the sum is over all
$\rta\in\RSys$ which restrict to $\rtl$ on $\ICSAp$. Consider their real
subspaces $\RSp{\rtl}=\cpl{\RSp{\rtl}}\cap\rLA$. It
follows from \req{plusRoot} that
\begin{equation} 
\elabel{plusRootR}
  \comm{\RSp{\rtl}}{\RSp{\rtm}}\neq 0 
\end{equation}
if $\rtl,\rtm$ and $\rtl+\rtm$ are restricted roots.

The restricted roots may be divided into positive and negative roots,
$\rRSys=\rRSysP\cup\rRSysN$, and among the positive restricted roots
a set of simple restricted roots $\srr{i}$, $i=0,...l$ is singled
out in the usual way.  They may be regarded as restrictions of the simple
roots $\sr{i}$ of $\RSys$.

Parabolic subgroups are conjugate to standard parabolic subgroups; what
is standard depends on the choice of $\ICSAp$ and the division into
positive and negative roots.  The standard parabolic subgroups
$\IParbl$ are classified by proper subsets $\Iroots$ of the set of
simple restricted roots $\srr{i}$. Their defining feature is that they
all contain the minimal standard parabolic subgroup $\IMinpb$ , which
corresponds to the empty set $\Iroots=0$.  Its Lie algebra $\Iminpb$
is as follows. One defines
$\Inilp[-]=\sum_{\rtl\in\rRSysN}\RSp{\rtl}$. This nilpotent subalgebra
appears as a summand in the Iwasawa decomposition of $\rLA$,
\begin{equation}
\elabel{Iwasawa}
  \rLA=\Ck+\ICSAp+\Inilp[-]
\end{equation}
One defines $\Ictz$ as the commutant of $\ICSAp$ in $\Ck$. Then the
minimal parabolic subalgebra is defined by
\begin{equation}
\elabel{minpb} 
  \Iminpb = \Ictz + \ICSAp + \Inilp[-] 
\end{equation}
and appears as a summand in the Bruhat decomposition
\begin{equation} 
\elabel{minBruhat}
  \rLA = \Inilp[+] + \Ictz + \ICSAp + \Inilp[-]\ ,
\end{equation}
with $\Inilp[+]=\sum_{\lambda \in \rRSysP}\RSp{\lambda}$.

\subsection{Gradings and Standard Parabolic Subalgebras}
 
It will be convenient for our purposes to introduce the other standard
parabolic subgroups together with gradations of $\rLA$.

First recall the following definitions (e.g.
\cite{Helgason:DGLGSS-78,FuchsSchwei:SLAR-97}): A \df{gradation} of a
Lie algebra $\rLA$ by an abelian semi-group $\Ggrp$ is a vector space
decomposition
\begin{equation}
\elabel{gradation}
  \rLA=\VSbigsum_{\Ginda\in\Ggrp}\Galg{\Ginda}
  \quad\text{ such that }\quad
  \comm{\Galg{\Ginda}}{\Galg{\Gindb}}=\Galg{\Ginda+\Gindb}\;.
\end{equation}
For symmetric spaces $\Z_2$-gradations play an important role. They
are in one-to-one correspondence with involutive authomorphisms of
$\rLA$. A $\Z$-gradation with the additional property
$\Galg{\Ginda}=\{0\}$ for $\absval{\Ginda}>n$ is called a
\df{$(2n+1)$-grading} if $n$ is the smallest such integer -- the
number of subspaces $\Galg{\Ginda}$ which are non-zero is $(2n+1)$.

For simplicity we assume temporarily that the zero vector is included
in $\rRSys$, and the root space $\RSp{(0)}$ corresponding to it is the
centralizer $\Ictz\LAsum\ICSAp$ of $\ICSAp$.

Given the subset $\srsubset$ of all simple restricted roots, a
$(2n+1)$-grading is obtained as follows.

Fix an ordering of the simple restricted roots such that
$\srsubset=\{\lambda_{m+1},\dots,\lambda_{\rk}\}$ while the remaining
simple restricted roots are $\lambda_{1},\dots,\lambda_{m}$. For a
restricted root $\lambda=\sum^\rk_{i=1}\srcoeff{i}\lambda_{i}$ define
the \df{level} to be
$\level[\srsubset]{\lambda}=\sum^m_{i=1}\srcoeff{i}$ which is an
integer.  A $(2n+1)$-grading of $\rLA$ is now defined by
\begin{equation}
\elabel{ngrading}
  \Galg{\Ginda}=\VSbigsum_{\level[\srsubset]{\lambda}=\Ginda}\RSp{\lambda}\;.
\end{equation}
From $\comm{\RSp{\rtl}}{\RSp{\rtm}}\subset\RSp{\rtl+\rtm}$ and
$\level[\srsubset]{\rtl+\rtm}=\level[\srsubset]{\rtl}+\level{\rtm}$ 
it is obvious that this
really defines an $(2n+1)$-grading for some $n$. 


Introducing the subalgebras
\begin{equation}
  \elabel{identNgrading}
  \InilpE{\pm}=\VSbigsum_{\Ginda=\pm 1}^{\pm n}\Galg{\Ginda}
  \;,\qquad
  \IleviE =\Galg{0}
  \;.
\end{equation}
the parabolic subalgebra is $\Iparbl[\srsubset]=\IleviE + \InilpE{-}$. 

We state a further decomposition. The socalled Levi subalgebra
$\IleviE$ contains the abelian subalgebra $\ICSApE$.  It is the
subspace of $\ICSAp$ on which all elements of $\Iroots$ are zero. The
Levi subalgebra $ \IleviE$ possesses an alternative characterization
as the commutant of $\ICSApE$ in $\rLA$. It may be further decomposed
as orthogonal sum
\begin{equation}
\elabel{}
  \IleviE = \ICSApE\LAsum\IctzI \ ,
\end{equation}
orthogonal with respect to the Killing form $\Killing{\cdot}{\cdot}$.
The root system of the complexification of $\IctzI$ consists of the
roots $\rta$ of $\rLA$ with $\level{r(\rta)}=0$, where $r$ is 
the restriction of roots.

In total, the parabolic subalgebra becomes
\begin{equation} 
\elabel{}
  \Iparbl=\IctzI\VSsum\ICSApE\VSsum\InilpE{-} 
\end{equation}
The Bruhat decomposition generalizes to 
$\rLA=\InilpE{+}\VSsum\IctzI\VSsum\ICSApE\VSsum\InilpE{-}$. 
For any $\srsubset$, a maximally non-compact Cartan subalgebra of $\IleviE$ 
is also a maximaly non-compact Cartan subalgebra of $\rLA$. 

\subsection{A Three-grading}

To prove \rst{threegrading} below, we need some properties of the
gradation \req{identNgrading}. We extend the definition of the level
function $\level[\srsubset]{\cdot}$ on restricted roots to roots by
$\level[\srsubset]{\rta}=\level[\srsubset]{r(\rta )}$ where $r$ is the restriction map.

\begin{lemma}[nonZ]
All the subspaces $\Galg{\Ginda} \neq 0$, for $\Ginda=-n...+n$
and 
$\comm{\Galg{\Ginda}}{\Galg{-1}}\neq 0 $ for $\Ginda = 1...n$
\end{lemma}

\begin{proof}
  For each root $\rta\in \RSys$ the root space $\RSp{\rta}$ is nonzero
  by definition.  For any positive root $\rta$ there is a simple root
  $\sr{i}$ such that $\rta - \sr{i}$ is a positive root or zero. Now
  there are two cases. If $\sr{i}\in\srsubset$, then
  $\level[\srsubset]{\rta -\sr{i}} = \level[\srsubset]{\alpha}$, and
  otherwise $\level[\srsubset]{\rta -\sr{i}} =
  \level[\srsubset]{\alpha} -1$. In the first case there is another
  simple root $\sr{j}$ such that $\rta - \sr{i} - \sr{j}$ is a simple root or
 zero, and the same
  two cases are possible for $\rta - \sr{i} - \sr{j}$. This procedure
  is repeated until a positive or zero root $\rtb$ is found such that
  \begin{equation}
  \elabel{} 
    \level[\srsubset]{\rtb} = \level[\srsubset]{\rta}-1 \qquad \mbox{and}
    \qquad \level[\srsubset]{\rtb + \sr{k}} = \level[\srsubset]{\rta}  
  \end{equation}
  where $\sr{k}$ is the last simple root found. It obeys
  $\level[\srsubset]{\sr{k}}=1$ This happens in a finite number of
  steps because the positive roots are linear combinations of simple
  roots with non-negative coefficients, and $\level[\srsubset]{0}=0$.
  Thus for each positive root $\rta$ there is such a root $\rtb$.
  
  Suppose $\RSp{\rta}\subset\Galg{\Ginda}$ with $\Ginda>0$.  Then, by
  \req{plusRoot}, $0\neq\RSp{\rtb} =
  \comm{\RSp{\rtb+\sr{k}}}{\RSp{\sr{k}}}\subset\comm{\Galg{\Ginda}}{\Galg
    {-1}}$. This proves the second statement of the lemma
  and shows at the same time that $\Galg{\Ginda}\neq 0$ implies
  $\Galg{\Ginda -1}\neq 0$, hence $\Galg{\Ginda}\neq 0 $ for $\Ginda =
  n...1$.
  With any root $\rta$, there is also a root $-\rta$ of level
  $-\level[\srsubset]{\rta}$. Therefore $\Galg{-\Ginda}\neq 0$ if
  $\Galg{\Ginda}\neq 0$. This completes the proof of the lemma.
\end{proof}

\begin{corr}
  In each restricted root system there is a unique \df{highest} root
  $\rrtmx$. It is defined by the property that $\rrtmx+\srr{i}$ is not
  a root for any simple restricted root $\srr{i}$. The coefficients
  $\CoxL{i}$ in the expansion $\rrtmx=\sum_{i=1}^\rk\CoxL{i}\srr{i}$
  are called \df{Coxeter labels}. 
\end{corr}

\begin{proposition}[threegrading]
  If $\rLA$ is simple\begin{corr}, $\Iroots$ non-trivial,\end{corr}%
  \rule{0pt}{0pt} and $\comm{\InilpE{+}}{\InilpE{-}}\subseteq\IleviE$
  then
  \begin{enumerate}
  \item the generalized Bruhat decomposition $\rLA= \InilpE{+} \oplus
    \IleviE \oplus \InilpE{-}$ is a three-grading of $\rLA$ with grading
    given by the level function $\level[\srsubset]{\cdot}$.
  \item $\InilpE{+}$ and $\InilpE{-}$ are abelian
  \item
    There is exactly one restricted root $\srrext$ not in $\srsubset$. Its Coxeter label is $1$. \\
    If the complexification $\cpl{\rLA}$ of $\rLA$ is simple, there is
    exactly one simple root $\srext$ of level
    $\level[\srsubset]{\srext}=1$; its restriction is $\srrext$
  \item
    $\ICSApE$ is 1-dimensional.
  \end{enumerate} 
\end{proposition}

\begin{proof}
1. The level function $\level[\srsubset]{\cdot}$ furnishes a
  $(2n+1)$-grading for some $n$. Since $\InilpE{\pm}$ are not empty,
  $n > 0$. Since $\Galg{\Ginda}\neq 0$ for $-n \leq \Ginda \leq n$,
  $n=1$ follows from $\Galg{2}=0$.
  
  Suppose $\Galg{2}\neq 0$. Then $\Galg{2}\subset \InilpE{+}$ and
  $0\neq \Galg{-1}\subset \InilpE{-}$, therefore
  $\comm{\InilpE{+}}{\InilpE{-}}\supset\comm{\Galg{2}}{\Galg{-1}}$. But
  $\comm{\Galg{2}}{\Galg{-1}}\neq 0 $ by \rst{nonZ}.  On the
  other hand, $\comm{\Galg{2}}{\Galg{-1}}\subset\Galg{1}$ by the
  defining property of a gradation. This contradicts the assumption
  $\comm{\InilpE{+}}{\InilpE{-}}\subset\IleviE=\Galg{0}$.
  \begin{corr}Therefore $\Galg{2}=0$ and $\rLA=\Galg{-1}\VSsum\Galg{0}\VSsum\Galg{1}$ is a three-grading.\end{corr}
  
2. By the three-grading property, $\comm{\Galg{1}}{\Galg{1}}=0$ and
  $\InilpE{+}=\Galg{1}$, implying that $\InilpE{+}$ is abelian, and
  similarly for $\InilpE{-}$.
  
3.As mentioned above, the Coxeter labels $\CoxL{i}$ are the coefficients in
  the decomposition of the highest restricted root in simple
  restricted roots $\srr{i}$, viz.
  $\rrtmx=\sum_{i=1}^l\CoxL{i}\srr{i}$. Order the restricted roots as
  described above, so that $\srr{1},\dots,\srr{m}$ are not in
  $\srsubset$.  Then $\level[\srsubset]{\rrtmx}=\sum_{i=1}^m
  \CoxL{i}$.  Since the restricted root system is irreducible, the
  Coxeter labels $\CoxL{i}$ are non-zero positive integers. Also,
  $m>0$ because $\srsubset$ is a proper subset of the set of all
  simple restricted roots. Therefore,
  $\absval{\level[\srsubset]{\rrtmx}}\leq 1$ is only possible when
  $m=1$ and $\srr{1}$ is the only simple restricted root not in
  $\srsubset$, and it has Coxeter label $\CoxL{1}=1$.
  
  If $\cpl{\rLA}$ is simple, it has a connected Dynkin diagram. It
  follows that $\rtb=\sum\sr{i}$ (sum over all simple roots of
  $\rLA_\C$) is a root.  (This follows from property ii) in the proof
  of \rst{contrlevel}).  If there were more than one simple root of
  non-zero level, $\rtb$ would have level greater one, contradicting
  the 3-grading.
  
4. As a real vector space, $\ICSAp$ is the dual of the real linear
  space spanned by the simple restricted roots. By 3), $\ICSApE\subset
  \ICSAp$ is the subspace on which all but one of the simple
  restricted roots vanish.  Therefore it is 1-dimensional.
\end{proof}

\begin{proof}[ of \rst{abelN1dimA}]
  \Rst{abelN1dimA} is merely a restatement of assertions 2. and 4. of
  the above proposition.
\end{proof}

\subsection{An Irreducible Representation}

Let $\aICSApE$ be the connected subgroup of $\rLG$ with Lie algebra 
$\ICSApE$.
For the analysis of the existence of an infinitesimal causal structure
in \rsec{infcausGCS} the following result is needed.

\com{For this statement and proof, have another look at PhD}

\begin{proposition}[prop:irr]
  If the complexification $\cpl{\rLA}$ of $\rLA$ is simple and
  $\comm{\InilpE{+}}{\InilpE{-}}\subseteq\IleviE$ then $\InilpE{+}$
  carries an irreducible representation of $\IctzI$ and elements of
  $\aICSApE$ act on $\InilpE{+}$ as multiplication by positive
  constants.
\end{proposition}
\begin{proof}
  \begin{del}The {\em level} of a root $\rta$ shall shall mean
    $\level[\srsubset]{\alpha}$.\end{del}
  
  $\InilpE{+}$ carries a representation of $\IleviE$ as a consequence
  of the 3-grading. Because $\IleviE$ is reductive, this
  representation is fully reducible.  By Schurs lemma, the assertion
  of the lemma is equivalent to the statement that this representation
  is irreducible. Elements of $\aICSApE$ act as multiplication with
  {\em real} constants because $\ad(X)$ is hermitean for elements
  $X\in \Cp$. The real constants are {\em positive} because $\aICSApE$
  is connected.
  
  \com{Im Folgenden muesste eigentlich zur Komplexifizierung
    uebergegangen werden. Vielleicht kann man das loesen mit dem
    folgenden Satz: For the following argument, it is necessary to use
    the complexifications of the various subalgebras. For simplicity
    we will assume this implicitly.}
  
  The maximally non-compact Cartan subalgebra $\ICSAk\LAsum\ICSAp$ of
  $\rLA$ is also a Cartan subalgebra of $\IleviE$.  The weights of the
  representation of $\IleviE$ on $\InilpE{+}$ are therefore the roots
  $\rta$ of $\rLA$ whose root spaces span $\InilpE{+}$, i.e.  the
  roots of level one.  (This can be seen more explicitly in
  \cite{deRiese:PhD-04}.) Because root spaces are one-dimensional, all
  these weights have multiplicity~one.
  \begin{del}
    The roots of level $1$ are positive because the unique simple root
    of level $>0$ has the positive coefficient 1.
  \end{del}
  
  Every irreducible subrepresentation has a lowest weight vector and
  is uniquely determined by it. Because the weights have multiplicity
  1, two different lowest weight vectors have different weight.
  Irreducibility is proven by showing that the lowest weight $\mu$ is
  unique. As a lowest weight, it has the property that $\mu-\sr{i}$
  is not a weight for any simple root $\sr{i}$ of $\IctzI$, and
  therefore not a root of $\rLA$. By definition, the roots $\sr{i}$
  of $\IctzI $ are the roots of $\rLA$ of level $0$.
  
  By \rst{threegrading}, there is a unique simple root
  $\srext$ of $\rLA$ of level 1.  It is a lowest weight by the
  properties of simple roots.  By \rst{contrlevel} below, any other
  root $\mu $ of level 1 can be obtained from $\srext$ by adding
  simple roots of $\rLA$ one by one, in such a way that a root is
  obtained at each stage.  Simple roots of level $>0$ cannot be added
  in this process, because there are no roots of level $\geq 2$, by
  the three-grading. Therefore the added roots must be roots $\sr{i}$
  of $\IctzI$.  It follows that $\mu -\sr{i}$ is a root for some
  simple root $\sr{i}$ of $\IctzI$, except for $\mu = \sr{e}$. This
  proves uniqueness of the lowest weight, hence irreducibility.
\end{proof}

\begin{lemma}[contrlevel]
  Let $\rta=\sum\rta^i\sr{i}$ be the expansion of a root $\rta$ into
  simple roots.  All positive roots $\rta$ in $\RSys$ with the property that
  the coefficient $\rta^k$ is non-zero can be constructed by
  successively adding simple roots starting with $\sr{k}$ such that
  every intermediate sum is a root.
\end{lemma}
\begin{proof}
  From the well known construction of the root systems
  \cite{Cornwell:GTP2-84,deRiese:PhD-04} it is known that each
  positive root can be obtained by successively adding simple roots
  such that each intermediate sum is also a positive root. It remains
  to be shown that this construction can be started with $\sr{k}$.

  For a positive root $\rta$ define $T_\rta$ to be the subdiagram of
  the Dynkin diagram of $\RSys$ consisting of the simple roots
  $\sr{j}$ for which $\rta^j\neq 0$ in the expansion. The following
  three facts are needed: 
  \vspace*{-1.5ex}
  \begin{itemize}
  \item[(i)] The subdiagram $T_\rta$ for a positive root $\rta$ is a
    connected Dynkin diagram.
  \item[(ii)] If $\rta$ is a positive root and $\sr{i}$ a simple root
    for which $\rta^i=0$ and $\sr{i}$ is connected to the subdiagram
    $T_\rta$ within the Dynkin diagram of $\RSys$ then $\rta+\sr{i}$
    is a positive root.
  \item[(iii)] Assume that in the construction a sequence like
    $(\rta+\sr{i})+\sr{k}$ appears, where $\rta$ has coefficients
    $\rta^i\neq 0$, $\rta^k=0$, and $\rta+\sr{i}$ is a root. Then
    $\rta+\sr{k}$ is also a root, and the construction may be reorderd
    $(\rta+\sr{k})+\sr{i}$.
  \end{itemize}
  \vspace*{-1.5ex}%
  We will first show that the statement of \rst{contrlevel} follows
  from (i), (ii), and (iii): When using (iii) repeatedly to reorder a
  given construction of some positive root $\rta$,
  in the resulting construction  the root $\rta'=\sum
  1\:\sr{i}$ appears, where summation is only over simple roots $\sr{i}$ in
  $T_\rta$. Because of (i) and (ii) this root $\rta'$ may be
  constructed starting from $\sr{k}$ and adding all the other simple
  roots in $T_\rta$ by moving along the connected graph.
  
  To show (i), (ii), and (iii) the following fact is needed:
  Given any two roots $\sr{i}$ and $\rta$ the \df{$\sr{i}$-string of roots
    through} $\rta$ is defined to be the set of all roots of the form
  $\rta + k \sr{i}$.  Theorem 13.5.IX in \cite{Cornwell:GTP2-84} states
  that $\rta+k\sr{i}$ is a root for all $-\RStrNeg\le k\le\RStrPos$ for
  some positive or zero integers $\RStrNeg$ and $\RStrPos$.  For the
  $\sr{i}$-string of roots through $\rta$ these integers are
  constrained by
  \begin{equation}
  \elabel{}
    \RStrPos = \RStrNeg - \CartanNum{\sr{i}}{\rta}\;.
  \end{equation}
where $\CartanNum{\sr{i}}{\rta}$ are Cartan integers. 
  Statement (ii) follows since $\RStrNeg=0$ and
  $\CartanNum{\sr{i}}{\rta}$ is negative in that situation. If on the
  other hand $\sr{i}$ is not connected to $T_\rta$,
  then $\CartanNum{\sr{i}}{\rta}$ is zero and consequently $\rta+\sr{i}$ is
  no root. This justifies (i). Statement (iii) can be seen because
  $\rta+\sr{k}$ is a root, which follows from (ii).
\end{proof}

\subsection{Space-time as a homogeneous space for $\aCk$}

For the analysis of global causal structure, another decomposition of
the Lie algebras of parabolic subgroups will be needed.

Remember that $\quotient{\aCk}{\Cent}$ is the maximal compact subgroup
of $\quotient{\rLG}{\Cent}$, where $\Cent$ is the center of $\rLG$
which is infinite for groups $\rLG$ which are of interest to us.

We begin by writing down the group version of the above Lie algebra
decompositions. Let $\aIctzI$, $\aICSApE$, and $\aInilpE{-}$ be the
connected subgroups of $\rLG$ whose Lie algebras are $\IctzI$,
$\ICSApE$, and $\InilpE{-}$, respectively.  Let $\ICtzIK$ be the
centralizer of $\ICSApE$ in $\aCk$.  With the definition
$\ICtzI=\ICtzIK\aIctzI$, the parabolic subgroup is
$\IParbl=\ICtzI\aICSApE\aInilpE{-}$.  By its definition, $\ICtzI$
contains the entire center $\Cent$ of $\rLG$.  In cases of interest to
us, it follows that $\ICtzI$ has infinitely many connected components,
while $\aIctzI$ is the identity component of $\ICtzI$. The identity
component $\Homstab=\aIparbl$ of the parabolic subgroup is 
\begin{equation} 
\elabel{H} 
  \Homstab=\aIparbl=\aIctzI\aICSApE\aInilpE{-}\ .
\end{equation}
We will be interested in a particular $\Iroots$.  The groups
introduced in the introduction are $\Dil=\aICSApE$, $\Spc=\aInilpE{-}$
and $\Lorentz=\aIctzI$ for this particular $\Iroots$, and $\aIctzI$ is
the generalized Lorentz group. It is connected, by its definition. Its
center is finite if it is semi-simple.

Consider the Iwasawa decomposition of the generalized Lorentz group
\begin{equation} 
\elabel{}
  \aIctzI= \aIctzIK\aICSApI\aInilpI{-}, \quad 
  \aIctzIK=\aIctzI\cap\aCk \ . 
\end{equation}

 For the conformal group $SO(2,4)$, $\aIctzIK$ is the
group of rotations, $\aICSApI$ consists of the Lorentz boosts in
$z$-direction, and $\aInilpI{-}$ is a two dimensional abelian subgroup
of the little group of a light-like vector pointing in $z$-direction.
It holds $\aICSAp=\aICSApI\aICSApE$,
$\aInilp[\pm]=\aInilpI{\pm}\aInilpE{\pm}$.

It follows from the definitions that the groups 
\begin{equation} 
\elabel{AN}
  \aICSAp=\aICSApI\aICSApE \mbox{ and } \aInilp[-]=\aInilpI{-}\aInilpE{-}
\end{equation}
agree with the subgroups $\aICSAp$ and $\aInilp[-]$ in the Iwasawa
decomposition $G=K\aICSAp\aInilp[-]$ of $\rLG$, whose Lie algebra
version is \req{Iwasawa}. Comparing with the definition \req{H} of
$\Homstab$, we find that
\begin{equation}
\elabel{SpacetimeComp}
  \Spacetime = \quotient{\rLG}{\Homstab} = \quotient{\aCk}{\aIctzIK}\ .
\end{equation}
This is the desired exposition of $\Spacetime$ as a homogeneous space for 
$\aCk$. It is noncompact because of the infinite center $\Cent$ of $\aCk$. 

Using this result, we obtain another realization of the tangent space
of $\Spacetime$ at $\Hid$.  Let $\IctzIK$ be the Lie algebra of
$\aIctzIK$.  Given a decomposition of $\Ck$ as a direct sum
$\Ck=\IctzIK\oplus\homrestk$,
\begin{equation}
\elabel{homrestk}
  \Tang{\Hid}{\Spacetime}=\quotient{\Ck}{\IctzIK} \simeq \homrestk \ ,
\end{equation}
Inserting the decomposition of $\Ck$ in the Iwasawa decomposition
\req{Iwasawa}, and noting the Lie algebra versions of decompositions
\req{AN}, we see that $\rLA= \homstab \oplus \homrestk$, where
$\homstab$ is the Lie algebra of $\Homstab=\aIparbl$. Therefore
$\homrestk$ can be used as a choice of $\homrest$ in \rst{repONq},
which asserts that $\homrestk$ carries a representation of $\Homstab$.
It is equivalent to the representation on $\trl$ and on
$\Tang{\Hid}{\Spacetime}$.  This will be used in the discussion of the
causal structure.

Comparing with the generalized Bruhat decomposition in
\rst{threegrading}, we see that a possible choice for $\homrestk$ is
\begin{equation} 
\elabel{}
  \homrestk = \left( \InilpI{-}\oplus\InilpI{+}\right)\cap \Ck \ .
\end{equation} 

\begin{lemma}[lemma:rotinv]
  If the generator $\I\energy$ of the center $\Ckc$ of $\Ck$ is not in
  $\IctzIK$ then $\InilpE{+}$ contains a vector which is
  invariant under elements of $\aIctzIK$ (``rotations'').
\end{lemma}
\begin{proof}
  $\InilpE{+}$ and $\homrestk$ carry equivalent representations of
  $\IleviE$. But $\homrestk$ can be chosen such that $\Ckc \subset
  \homrestk$ if the generator $\I\energy$ of $\Ckc$ is not in
  $\IctzIK$. It commutes with $\IctzIK$ because it is in the center of
  $\Ck$. By \rst{repONq}, it therefore defines a $\aIctzIK$-invariant
  vector in $\homrestk$.
\end{proof}
%

\section{Causal Structure}
\label{sec:ICS}

The global causal structure needed can be described infinitesimally by
a specification of tangent vectors which non-spacelike curves are
allowed to have. These lie in cones which are non-trivial in a certain
way described below and invariant under the assumed symmetry. We will first
 develop criteria for the existence of such an
infinitesimal causal structure.

\subsection{Infinitesimal Causal Structure}
\label{sec:infcausGCS}

We will summarize a few definitions and facts about cones and
cone-fields in homogenous spaces
\cite{Segal:MCEA-76,FarautOlafsson:CSSSGHA-95}

A \df{cone} $\cone$ is a subset of a vector space which contains
$\lambda v$ for all $v\in\cone$ and $\lambda>0$.  A cone is \df{proper} if
$C\cap -C=\{0\}$ and \df{generating} if its linear span is $V$.  A
convex, proper, generating, and closed cone is called \df{regular} cone.

An \df{infinitesimal causal structure} or \df{cone-field} $\conefield$
on a $\Dim$-dimensional manifold $\Spacetime$ is an assignment of a
regular cone $\conefield(x)$ in $\Tang{x}{\Spacetime}$ to each
$x\in\Spacetime$ such that there are an open covering $\{U_i\}_{i\in
  I}$ of $\Spacetime$, a cone $\cone\in\R^\Dim$, and (smooth or
analytic) maps
$\varphi_i:U_i\times\R^\Dim\rightarrow\Tang{}{\Spacetime}$ with
$\varphi_i(x,\cone)=\conefield(x)$.

As before, let the Lie group $\rLG$ act differentiably on $\Spacetime$ and 
call the
action $\Homact$.  Let the derivative of $\Homact(g)$ at $x$ be
$\homact^x(g)= \Tder{x}{(\Homact(g))}:\Tang{x}{\Spacetime} \rightarrow
\Tang{\Homact(g)x}{\Spacetime}$.

A causal structure is \df{$\rLG$-invariant} if for each $g\in\rLG$ and
$x\in\Spacetime$ the derivative $\homact^x(g)$ maps the cone-field into
a itself:
\begin{equation}
\elabel{invconefield}
\homact^x(g)\conefield(x)\subset\conefield(\Homact(g)x)\,.
\end{equation}

If the action of $\rLG$ is transitive, so that
$\Spacetime=\quotient{\rLG}{\Homstab}$, there is a bijection from the
set of regular $\Homstab$-invariant cones in $\Tang{\Hid}{\Spacetime}$
with $\Hid=\coset[\Homstab]{\Gid}$ onto the set of $\rLG$-invariant
infinitesimal causal structures on $\Spacetime$
(\cite{FarautOlafsson:CSSSGHA-95}):
\begin{equation}
  \elabel{bijConeConefield}
  \cone\mapsto(\conefield:\coset[\Homstab]{g}\mapsto\homact^\Hid(g)\cone)\,.
\end{equation}
Therefore the action $\homact^\Hid$ of $\Homstab$ on
$\Tang{\Hid}{\Spacetime}$ and the cones invariant under it have to be
studied. We will drop the ``$\Hid$'' in the notation were appropriate:
$\homact(g)=\homact^\Hid(g)$.

If $\Homstab$ is trivial, i.e. if $\Spacetime\isom\rLG$, then the
$\Homstab$-invariant regular cones in $\Tang{\Gid}{\rLG}$ are
just all regular cones. Therefore, there always exist
$\rLG$-invariant infinitesimal causal structures on $\rLG$.

Some general results on invariant cones in arbitrary vector spaces
follow. They can be found in
\cite{Paneitz:ICCCSLAG-81,Vinberg:ICCOLG-82,Segal:MCEA-76} in a
slightly different form.

Let $\mLG$ be any connected semi-simple real Lie group and $V$ a real
finite-dimensional $\mLG$-module. Define $\mCk$ to be a maximal
subgroup of $\mLG$ such that the image in $\Gl{V}$ is compact.  Note
that this is not necessarily the compact factor of a Cartan
decomposition of $\mLG$. This notation is fixed for the following
three lemmata.

\begin{lemma}[coneKstableV]
  There is a $\mCk$-stable $v_\mCk\in V$, $v_\mCk\neq 0$ if and only
  if there exists a $\mLG$-invariant closed convex cone $\cone$ in $V$
  with $\cone\neq\cone\cap-\cone$. In this case there will always be a
  proper cone of this type.
\end{lemma}
Let us explicitly state that the lemma holds true if $V=\R$ and $M$ and $U$ 
are trivial. In this case, the half lines $\R_\pm$ are invariant cones.
\begin{proof} 
  \cite{Paneitz:ICCCSLAG-81} For simplicity set $\mLG$ and $\mCk$
  equal to their images in $\Gl{V}$.  Let $\cone$ be a closed convex
  cone in $V$.  The following is a standard result about convexity 
(\$17.1 and \$16.3 in \cite{Koethe:TLRI-66}).
 Because  $\cone\neq\cone\cap-\cone$ there exists a
  linear functional $f$ in $V$ which is non-negative and non-trivial
  on $\cone$. 
  Select $v\in\cone$ with $f(v)>0$, then
  \begin{equation}\elabel{}
    v_\mCk=\int_\mCk k\cdot v \;\Haar{k}
  \end{equation}
  is $\mCk$-invariant and $f(v_\mCk)>0$. It follows from linearity that
  $v_\mCk\neq0$.
  
  Let on the other hand $v_\mCk$ be $\mCk$-invariant. Choose a scalar
  product $\HProd{}{}$ in the complexification $\cpl{V}$ of $V$ which
  is left invariant by the compact form of the complexification
  $\cpl{\mLG}$ of $\mLG$. Let $\mLA=\Cm\oplus\Cp$ be a Cartan
  decomposition of the Lie algebra $\mLA$ of $M$. $X\in\Cp$ is hermitian and $\Gexpsm{X}$
  positive-definite hermitian. Because of the invariance of $v_\mCk$
  and the Cartan decomposition $\mLG=\mCp\mCk$, the $\mLG$-orbit of
  $v_\mCk$ is equal to the $\mCp$-orbit. For $u,w$ in this orbit we
  have $\HProd{u}{w}=\HProd{\Gexpsm{X}v_\mCk}{\Gexpsm{Y}v_\mCk}
  =\HProd{\Gexpsm{Y}\Gexpsm{X}v_\mCk}{v_\mCk}$. With the Cartan
  decomposition $\Gexpsm{Y}\Gexpsm{X}=\Gexpsm{Z}k$, $k\in \mCk$ and the
  positive-definiteness of $\Gexpsm{Z}$ it is $\HProd{u}{w}>0$.  This
  extends by linearity to the convex cone generated by this orbit. This
  convex cone is $\mLG$-invariant, and by continuity we have
  $\HProd{u}{w}\ge0$ for $u,w$ in the closure $\cone$. If $v$ and $-v$
  are in $\cone$, $\HProd{v}{-v}\ge0$ i.e.  $v=0$, thus $\cone$ is
  proper.
\end{proof}

Irreducibility will be needed for the cones to be regular:

\begin{lemma}[irrCreg]
  If ~$V$ is an irreducible $M$-module and contains a non-trivial
  $M$-invariant convex closed cone $\cone$, $\cone$ is regular.
\end{lemma}
\begin{proof}
  Both the linear span of $\cone$ and the subspace $\cone\cap -\cone$ are
  invariant linear subspaces of $V$ which can only be $V$ or $\{0\}$.
  If the linear span of $\cone$ is $\{0\}$ then $\cone$ is trivial. If $\cone\cap
  -\cone$ is $V$ then $\cone$ is $V$.
\end{proof}

The irreducibility will be important for global causality, because the
following result of \cite{Vinberg:ICCOLG-82} will be needed:

\begin{lemma}[irrVuniq]
  If ~$V$ is irreducible and contains a non-zero $\mCk$-stable vector
  $v_\mCk$ then $v_\mCk$ is unique up to scalar multiplication with
  respect to this property.
\end{lemma}

\subsection{Generalized Conformal Symmetry}
\label{infcausGCS}

Referring to our standing assumptions 1 and 2 we will now specialize
to the case where $\Homstab$ is a parabolic subgroup of $\rLA$, and
the associated generalized Bruhat decomposition \req{Bruhat} satisfies
eq. \req{assumption:1}. We are interested in simple $\rLA$, and we may
assume that $\cpl{\rLA}$ is simple, because this will be implied for
Lie algebras whose maximal compact subalgebra has nontrivial center.

We wish to apply the results of the previous subsection to the action
of the generalized Lorentz group $\mLG=\aIctzI$ on
$V=\InilpE{+}\simeq \Tang{\Hid}{\Spacetime}$
%
%
where $\srsubset $ are
particular subsets of the set of restricted roots such that the
hypotheses of \rst{prop:irr} are fulfilled.  Therefore the
result of this proposition becomes available, and we may conclude that
any $\Lorentz$-invariant cone in $V$ will also be invariant under $\Homstab=\Lorentz\Dil\Spc$,
because $\Spc$ acts trivially by hypothesis, and elements of $\Dil$ act by
multiplication with positive constants, and such multiplication
carries any cone into itself by definition.

Using \rst{coneKstableV} and \rst{irrCreg} to ensure
$\Lorentz$-invariance, we obtain the following corollary of
\rst{prop:irr}.

\begin{corollary}[infcausalstruct]
  Under the same hypotheses, the following holds true. If $\Lorentz$
  is semi-simple there is a $\Lorentz$-invariant regular cone in
  $\InilpE{+}$ if and only if there is a $\mCk$-stable vector, where
  $\mCk=\Lorentz\cap\aCk=\aIctzIK$.  If and only if this is the case,
  there exists an $\rLG$-invariant infinitesimal causal structure on
  $\Spacetime$.
\end{corollary}

\begin{proof}[ of  first part of \rst{theo:4}]
  This follows from \rst{infcausalstruct} and theorem \ref{theo:1} which
  asserts that $\InilpE{+}\simeq \Tang{\Hid}{\Spacetime}$ for 
$\Iroots $ such that $  \Homstab=\aIctzI\aICSApE\aInilpE{-}$.
\end{proof}

The question arises what is the $\mCk$-stable vector and
whether this infinitesimal causal structure is a 
 global causal structure in one of
the senses defined in \cite{HawkingEllis:LSSST-73}. In
the next section it will be shown that 
the diffeomorphism $  \Spacetime=\Cauchy\times \R$ which is required by 
global hyperbolicity will require that $\Ck$ has nontrivial center $\Ckc$ not contained in the Lie algebra $\lorentz$ of $\Lorentz$ and the  Lie subgroup 
of $\rLG$ associated with $\Ckc$ must be isomorphic to $\R$. Conversely,
this property yields a $\mCk$-stable vector and implies global hyperbolicity.

\subsection{Global Causality}
\label{sec:globalcausality}

We recall a few definitions: A differentiable curve
$\map{\crve}{I}{\Spacetime}$, where $I\subset\R$ is an interval, is
(future-directed) \df{non-spacelike} with respect to an infinitesimal
causal structure $\conefield$ if the tangent vector at $\crve(\tau)$
is in the cone $\conefield(\crve(\tau))\setminus\{0\}$ for all
$\tau\in I$. For $p\in\Spacetime$ the \df{causal future} resp.
\df{past} $\gcone[\pm]{p}$ of $p$ is the set of points
$q\in\Spacetime$ with a non-spacelike curve $\crve$ such that
$\crve(0)=p$ and $\crve(\pm\tau)=q$ with $\tau\ge0$.

An infinitesimal causal structure $p\mapsto\conefield(p)$ is
\df{strongly causal}, if every neighbourhood of a point
$p\in\Spacetime$ contains a neighbourhood of $p$ which no
non-spacelike curve intersects more than once, i.e. the neighbourhood
is mapped to an interval by $\inv{\crve}$. An infinitesimal causal
structure is \df{globally hyperbolic} if it is strongly causal and if
for $p,q\in\Spacetime$ the set $\gcone{p}\cap\gcone[-]{q}$ is compact.
A homogeneous space $\Spacetime $ of $\rLG$ will be called globally hyperbolic
if it admits a $\rLG$-invariant infinitesimal causal structure which is globally hyperbolic.

A set $S\in\Spacetime$ is called \df{acausal}, if for $p\in S$ no
$q\in S$, $p\neq q$ is in $\gcone{p}$.

The \df{edge} of an acausal set $S\in\Spacetime$ is the set of all
$p\in\closure{S}$ such that in every neighbourhood of $p$ there are points
$q$ in the future and $r$ in the past of $p$ which can be joined by a
timelike curve within the neighbourhood without intersecting $S$. If
$S$ is a submanifold without boundary, the edge is empty, because the
boundary clearly contains the edge. 

A set $\Cauchy\in\Spacetime$ is called a \df{Cauchy surface} for
$\Spacetime$, if it is acausal, has no edge, and each inextendible
non-spacelike curve in $\Spacetime$ intersects $\Cauchy$.  A criterion
for global hyperbolicity is given by the

\begin{proposition} \label{proposition3}
  \cite{HawkingEllis:LSSST-73} $\Spacetime$ is globally hyperbolic if and only
  if it has a Cauchy surface. Then $\Spacetime$ is homeomorphic to
  $\R\times\Cauchy$, where $\Cauchy$ is a manifold of codimension 1
  and each $\{a\}\times\Cauchy$, $a\in\R$ is a Cauchy surface for
  $\Spacetime$.
\end{proposition}
 
%

Now we turn to sufficient conditions for the existence of a global
causal structure. The proof of the only if part of part ii) of \rst{theo:4}
 will follow from \rst{requirements} below by showing that its
hypothesis are fulfilled. To prepare for \rst{requirements} we formulate and prove three lemmas.

\begin{lemma}[Rcrosssigma]
\label{Rcrosssigma}
\begin{equation}
\Spacetime \simeq \R\times\Cauchy
\end{equation}
if and only if $\Ck$ has nontrivial center $\Ckc$ not contained in the Lie algebra $\lorentz$ of $\Lorentz = \aIctzIK$ and the connected Lie subgroup 
$\aCkc$ of $\rLG$ with Lie algebra $\Ckc$ is isomorphic to $\R$.

In this case, $\aCk=\aCks\times\aCkc$ and 
\begin{equation}
  \Spacetime=\Cauchy\times\aCkc ,
 \qquad \Cauchy = \quotient{\aCks}{\aIctzIK}\, \label{CauchyFact} .
\end{equation}
\end{lemma}
\begin{proof}
Consider the homogenous space $\Spacetime=\quotient{\aCk}{\aIctzIK}$
from equation \req{SpacetimeComp}. If $\aCk$ is compact $\Spacetime$
cannot have the topology described above.  However, in a Cartan
decomposition of a simple Lie group $\rLG=\aCk\aCp$ the ``compact''
factor $\aCk$ is non-compact or has a noncompact covering if and only
if the Lie algebra $\Ck$ of $\aCk$ has a non-trivial center $\Ckc$.
\begin{artonly}
  Using several facts which can be found in \cite{Helgason:DGLGSS-78}
  this can be seen as follows: Generally, $\Ck$ is reductive
  \todo{Prop.  6.6 Ch. 2} and $\Ckc$ is at most one-dimensional%
  \todo{From classification of Cartan involution}. I.e. we have
  $\Ck=\Cks\oplus\Ckc$ where $\Cks$ is semi-simple. The connected Lie
  subgroup $\aCks$ of $\rLG$ corresponding to $\Cks$ has to be compact%
  \todo{Th. 6.9 Ch. 2}.  However, the connected Lie subgroup $\aCkc$
  corresponding to $\Ckc$ may be non-compact and its universal covering is
 isomorphic to  $\R$.
\end{artonly}
If $\Ckc $ were contained in the Lie algebra $\lorentz$ of $\Lorentz$ then 
the noncompact factor $\aCkc$ would cancel out in 
$\quotient{\aCk}{\aIctzIK}$ and $\aCk$ would be compact. 
 \end{proof}

Given eq.(\ref{CauchyFact}) the next task is to show that the 
quotient
$\Cauchy=\quotient{\aCks}{\aIctzIK}$ is a Cauchy surface.This is addressed 
by the following lemma. 
\newcommand{\timecomp}{t}
\newcommand{\spacecomp}{t^\perp}
\begin{lemma}\label{lemma10}
Assume the existence of an infinitesimal causal structure with cones 
$\conefield(p)$,  
  let $\timecomp:\Cauchy\times\aCkc\rightarrow\aCkc$ be the component
  map.  For any $a\in\aCkc\isom\R$ the submanifold $\Cauchy\times\{a\}$ is
  a Cauchy surface if the derivative of $\timecomp$ along each vector
  in the cone $\conefield(p)$ for all $p\in\Spacetime$ is strictly positive (or
  negative for all vectors).
\end{lemma}
\begin{proof}
  Let $\crve:I\rightarrow\Spacetime$ be any non-spacelike curve and
  choose any $\tau\in I$. Because the tangent vector
  $\Tder{\tau}{\crve}(1)$ of $\crve$ at $\tau$ lies in
  the cone $\conefield(\crve(\tau))$, it follows by the assumption in the lemma that
  $\der{\tau}{\timecomp\circ\crve(\tau)}>0$ and thus $\timecomp\circ\crve$
  is a strongly monotonous function.

    Thus, $\crve$ has exactly one intersection with
    $\Cauchy\times\{\timecomp\circ\crve(\tau)\}$. Since this is true
    for all non-spacelike curves, any $\Cauchy\times\{a\}$ is acausal.
    It has no edge because it is a submanifold. 

  \begin{dissonly}
    Thus, $\crve$ has exactly one intersection with
    $\Cauchy\times\{\timecomp\circ\crve(\tau)\}$. Since this is true
    for all non-spacelike curves, any $\Cauchy\times\{a\}$ is acausal.
  
    It also follows that $\Cauchy\times\{a\}$ has no edge: For some
    $p\in\Cauchy\times\{a\}$ let any two points $q$ and $r$ be in the
    future and past of $p$, respectively. The points $q$ and $r$ are
    in some submanifolds $\Cauchy\times\{a^{q}\}$ and
    $\Cauchy\times\{a^{r}\}$, respectively, for some $a^r$ and $a^q$
    with $a^r<a<a^q$. For any timelike curve $\crve$ connecting $q$
    and $r$ the function $\timecomp\circ\crve$ takes on the values
    $a^r$ and $a^p$ and because of the intermediate value theorem also
    $a$, i.e. $\crve$ intersects $\Cauchy\times\{a\}$, which therefore
    has no edge.
  \end{dissonly}  
  
  An inextendible non-spacelike curve $\crve$ has to intersect each
  $\Cauchy\times\{a\}$ because otherwise it would approach one
  $\Cauchy\times\{a\}$ arbitrarily closely and not have a limit point.
  This would mean that its tangent vector would approach the tangent
  bundel of $\Cauchy\times\{a\}$. \todo{Maybe this should be made more
    explicit.} Since the $\conefield(p)$ are closed cones which
  contain this vector, the tangent space of $\Cauchy\times\{a\}$ would
  actually intersect some $\conefield(p)$. A curve with tangent
  vectors in this intersection would clearly have zero derivative of
  $\timecomp$. A contradiction.
\end{proof}
Suppose that 
 $\Ck=\Cks\LAsum\Ckc$ (sum of ideals) and
$\IctzIK\subset\Cks$.  The suitably chosen subspace $\homrestk$ 
introduced before \req{homrestk}, which was defined to satisfy
$\Ck=\IctzIK\VSsum\homrestk$ may then be chosen to contain $\Ckc$, so that
$\Cks=\IctzIK\VSsum\homrests$, and $\homrestk=\homrests\oplus\Ckc$.

We finally end up with the following Lie algebraic criterion:
\begin{lemma}[coneintersection]
  $\Spacetime$ carries a $\rLG$-invariant globally hyperbolic causal
  structure, if there is an $\Homstab$-invariant regular cone
  $\cone$ in $\homrestk$ which has intersection $\{0\}$~with~$\homrests$.
\end{lemma}
\begin{proof}
  Since $\homrests$ has codimension $1$ in $\homrestk$, $\cone$ lies
  in one halfspace with boundary~$\homrests$.  The decomposition
  $\homrestk=\homrests\oplus\Ckc$ corresponds via the equivalence
  established in \rst{repONq} to the decomposition of the tangent
  space
  $\Tang{\Hid}{\Spacetime}=\Tang{\Hid}{\Cauchy}\VSsum\Tang{\Hid}{\aCkc}$.
  
From $\cone$ we obtain a $\rLG$-invariant infinitesimal causal ordering of 
$\Spacetime$ via	
 the bijection \req{bijConeConefield} between
  $\Homstab$-invariant regular cones in $\Tang{\Hid}{\Spacetime}$ and
  the infinitesimal causal structures in $\Spacetime$.  For
  $\Spacetime\ni p=\coset[]{g}$ we can choose the representative $g$
  to be in $\aCk$. Then $\homact(g)$ maps $\Tang{\Hid}{\aCkc}$ to
  $\Tang{p}{\aCkc}$ and $\Tang{\Hid}{\Cauchy}$ to $\Tang{p}{\Cauchy}$.
  $\cone$ is now mapped to a regular cone in a halfspace of
  $\Tang{p}{\Spacetime}$ with boundary $\Tang{p}{\Cauchy}$ which
  intersects $\Tang{p}{\Cauchy}$ only at zero. Identify $\aCkc$ -- and
  with it all $\Tang{p}{\aCkc}$ -- with $\R$ such that the
  $\Tang{p}{\aCkc}$-component of all elements in $\homact(g)(\cone)$
  is positive via this identification. This is then by continuity and
  connectedness of $\rLG$ the case for all $\homact(g)(\cone)$ for
  $g\in\aCk$.  The derivative of the function $\timecomp$ in direction
  of a vector with positive $\Tang{p}{\aCkc}$-component is clearly
  positive.
Therefore the hypothesis of Lemma \ref{lemma10} is satisfied and $\Cauchy$ is 
a Cauchy surface. By proposition \ref{proposition3} this implies global hyperbolicity.
\end{proof}

We are now in a position to formulate the criteria for the existence of a
$\rLG$-invariant infinitesimal causal structure on $\Spacetime$ that
turns $\Spacetime$ into a globally hyperbolic manifold:

\begin{theorem}[requirements]
\label{requirements}
  Let $\rLG$ be a connected simple Lie group and $\aIparbl[\Iroots]$
  the identity component of a parabolic subgroup satisfying
  assumptions 1 and 2. 
  $\Spacetime=\quotient{\rLG}{\aIparbl[\Iroots]}$ carries an
  infinitesimal causal structure such that $\Spacetime$ is globally
  hyperbolic if
  \begin{itemize}
  \item[(i)]   $\Ck$ has non-trivial center $\Ckc$ and $\aCkc$ is non-compact,
  \item[(ii)]  $\Ckc$ is not contained in $\IctzIK$, and
  \item[(iii)]   $\IctzI$ is semi-simple.
  \end{itemize}
  If $\Spacetime$ is globally hyperbolic (i) and (ii) always hold.
\end{theorem}
\begin{proof}
First we show that (i) and (ii) imply that $\Spacetime$ carries an
 infinitesimal causal structure.
  
  The conditions (i) and (ii) also lead to the existence of a
  $\aIctzIK$-stable vector in $\homrestk=\homrests\oplus\Ckc$: Any
  $\I\energy\in\{0\}\oplus\Ckc$ is such a vector.  Fix $\I\energy$ in the
  following and choose the identification of $\aCkc$ with $\R$ such
  that the derivative of $\timecomp$ in direction of $\I\energy$ is
  positive.
  
  With (iii) we know from \rst{infcausalstruct} that an $\aIctzI$-invariant
  regular cone exists in $\InilpE{+}$ and therefore there also exists
  an invariant cone $\cone$ in the equivalent module $\homrestk$.
  \Rst{AssumpSimple} and \rst{prop:irr} show that it is invariant
  under the action of $\aIparbl$.
$\cone $ determines a $\rLG$-invariant infinitesimal causal structure on 
$\Spacetime$ via the bijection \req{bijConeConefield}.
  
Next we use 
  \Rst{coneintersection} to show that the infinitesimal causal structure
  obtained from $\cone$ turns $\Spacetime$ into a globally hyperbolic
  manifold: If the intersection of $\homrests$ and $\cone$ were not
  $\{0\}$, this intersection would be a $\aIctzI$-invariant regular
  cone in $\homrests$. With (iii) and \rst{coneKstableV} there would be
  an $\aIctzIK$-stable non-zero vector in $\homrests$
  -- clearly not proportional to $\I\energy$. This contradicts
  \rst{irrVuniq}.

Finally, the last assertion of \rst{requirements} follows from Lemma
\ref{Rcrosssigma}. 
\end{proof}

\begin{proof}[ part ii) of \rst{theo:4}]

The {\em only if} part is Lemma \ref{Rcrosssigma}. 

{\em if part}: 
we must show that the hypotheses i), ii) and iii) of \rst{requirements}
are true. i) is true by hypothesis, and so is ii)
 (Actually ii) follows from i) as we shall see in a moment).

The proof of (iii) will rely on the classification 
 which will be performed in the next section. It will
rely on assumptions 1 and 2 and condition (i) in \rst{requirements}.
It turns out that for all cases $\aIctzI$ will be simple (zero in
the well known case of $\rLG=Sl(2,\R)$), and that $\I\energy$ will lie
outside of $\IctzIK$.
It will also follow from this classification that condition ii) of 
\rst{requirements} follows from assumptions 1 and 2 and from condition i). 

  Since condition (i) is equivalent to the
existence of positive energy representations, it will be a complete
classification.
\end{proof}

\section{Classification}

Non-compact real forms of simple complex Lie algebras can be
classified by Satake diagrams. This particular classification is
useful for us, because it makes reference to the maximally non-compact
Cartan subalgebra $\ICSAp\LAsum\ICSAk$ which in turn is the basis for
the classification of parabolic subgroups.

Recall the definition of the restricted root system $\rRSys$ in
section \ref{sec:parabolic}.  It is the set of non-zero restrictions
of the roots in the root system $\RSys$ of $\rLA$ to $\ICSAp$.  The
number of roots in $\RSys$ which is projected on $\rtl\in\rRSys$ is
called the multiplicity $\multipl{\rtl}$ of $\rtl$.  $\rRSys$ may be
described by an ordinary Dynkin diagram with additional information
about the multiplicities of $\srr{i}$ and $2\srr{i}$ for each simple
restricted root $\srr{i}$. Let $\srr{i}$ and $\srr{j}$ be simple
restricted roots such that $2\srr{i}$ is a root while $2\srr{j}$ is
not. Then these simple restriced roots will be denoted by
\begin{align}\elabel{}
  \DynDiaged{\DynR{--}&
    \DynOO{\raisebox{-1.15\height}{\shortstack{$\multipl{\srr{i}}$\\$\multipl{2\srr{i}}$}}}
    \DynR{--}&} &\text{ ~and }
  \DynDiaged{\DynR{--}&\DynO{$\multipl{\srr{j}}$}\DynR{--}&}
\end{align}

The real Lie algebra is completely described by the \df{Satake
  diagram}. It is the Dynkin diagram of $\RSys$ with two additional
elements:
\begin{description}
\item[Type 1] Simple roots in $\RSys$ which are restricted to zero in $\rRSys$
  are denoted by a filled node in the Satake diagram:%
  \DynText{\DynR{--}&\DynR{-}&\DynN{}\DynR{-}&\DynR{--}&}
\item[Type 2] Two simple roots in $\RSys$ which are restricted to the same
  element in $\rRSys$ are connected by an arrow:
  \raisebox{-1em}{\strut}
  \DynText{\DynR{-}&\DynO{}\DynR{-}\DynSym{rrr}{1}&\DynR{--}&\DynR{-}&\DynO{}\DynR{-}&}
\item[Type 3] All simple roots which are not of type 1 or type 2 are denoted by a regular node:%
    \DynText{\DynR{--}&\DynR{-}&\DynO{}\DynR{-}&\DynR{--}&}
\end{description}
The Satake diagrams for the simple real Lie algebras with positive
energy representations mentioned in \rsec{IntroPERep} are listed in
\rfig{SatakeDiag}. %
They are shown together with the Dynkin diagrams of the restricted
root systems. The Coxeter labels of the simple restricted roots are
indicated above the corresponding nodes.

\com{Das folgende stimmt nicht ganz. Ich habe es nur unabhaengig
  machen wollen. }
\begin{del}
  Unfortunately, a list of correspondences between Satake diagrams and
  other classifications of real Lie algebras could not be found in the
  literature.
To establish the correspondence, the Killing indices
$\KillingIndex$ are computed,
\begin{equation}
  \elabel{KillingIndex}
  \KillingIndex=\dim\Cp - \dim\Ck\;,
\end{equation}
i.e. the difference between the non-compact and the compact
dimensions. For the exceptional groups it is customary to include it
into the notation. For instance, $E_{7(-25)}$ has Killing index $-25$. 
\end{del}

\clearpage
\begin{mytable}{3}{&&}{|l@{\hspace*{0.5em}}|@{\hspace*{0em}}l|@{\hspace*{0em}}l|}%
                      {Type&\hfill$\RSys$\hfill\strut&$\hfill\rRSys$\hfill\strut}
                                                  \raisebox{2em}{\strut}%
$\mfrak{su}(m,n)$ \hspace{-4em}\raisebox{-0.9em}[0pt][0pt]{$_{(n=r, m+n=l+1)}$}
&\rule{0pt}{2em}\DynTab{\DynO{$1$}\DynR{--}\DynSym{rrrrrr}{-1}  &\DynO{$r$}\DynR{-}\DynSym{rrrr}{-0.5}  
                                           &\DynN{}\DynDDR & &\DynN{}\DynR{-} &\DynO{}\DynR{--} &\DynO{}}
&\DynTab{\DynOD{$2$}{$2$}\DynDDR             & &\DynOD{$2$}{$2$}\DynDR{=} 
            &\DynOOD{\raisebox{0pt}[\height][0pt]{${2l\!-\!4r\!+\!2}\hspace{-3.1em}\raisebox{-0.9em}{1}\hspace{0.7em}$}}{$2$}
}
                                   \hspace*{2.5em}\raisebox{-3em}{\strut}
\END
%
$\mfrak{su(m,m)}$ \hspace{-1.5em}\raisebox{-0.4em}[0pt][0pt]{$_{(m=\tfrac{l+1}{2})}$}
&\DynTab{\DynO{$1$}\DynDDR\DynSym{rrrrrr}{-1} & &\DynO{$\tfrac{l-1}{2}$}\DynR{-}\DynSym{rr}{-0.5}  
                                           &\DynO{$\tfrac{l+1}{2}$}\DynR{-} &\DynO{}\DynDDR & &\DynO{}}
&\DynTab{\DynOD{$2$}{$2$}\DynDDR             & &\DynOD{$2$}{$2$}
            &\DynOD{$1$}{$1$}\DynDL{=}}
\END
%
\Hline
                                                  \raisebox{1.5em}{\strut}%
$\mfrak{so}(2,\SOdim)$
&\DynTab{\DynN{} &\DynN{}\DynDL{=}\DynDDR &&\DynN{}\DynR{-} &\DynO{$1$}\DynR{-} &\DynO{$2$}}
&\DynTab{&\DynOD{\raisebox{0pt}[\height][0pt]{$2l\!-\!3\hspace*{1.3em}$}}{$2$}
         &\DynOD{$1$}{$1$}\DynDL{=}}
\END
%
\Hline
                                                  \raisebox{1.5em}{\strut}%
$\mfrak{sp}(\rk,\R)$
&\DynTab{\DynO{$1$}\DynDR{=} &\DynO{$2$}\DynR{-}  &\DynR{--} &\DynR{-} &\DynO{}\DynR{-} &\DynO{$\rk$}}
&\DynTab{\DynOD{$1$}{$1$}\DynDR{=} &\DynOD{$1$}{$2$}\DynDDR & &\DynOD{$1$}{$2$}}
\END
%
%
\Hline
%
%
$\mfrak{so}(2,\SOdim)$
&\rule{0pt}{3em}\DynTab{                    &\DynNR{}\DynD{-}\\
           \DynN{}\DynR{-}  &\DynN{}\DynDDR & &\DynN{}\DynR{-} 
           &\DynO{$1$}\DynR{-} &\DynO{$2$}}
&\DynTab{&\DynOD{\raisebox{0pt}[\height][0pt]{$2(l\!-\!2)\hspace*{1em}$}}{$2$}  &\DynOD{$1$}{$1$}\DynDL{=}}
\END
%
$\mfrak{so}^*(2\rk)$
&\rule{0pt}{3em}\DynTab{                    &\DynOR{$1$}\DynD{-}\\
           \DynN{}\DynR{-}  &\DynO{$2$}\DynR{-} &\DynN{}\DynDDR &
            &\DynN{}\DynR{-}&\DynO{$l/2$}\DynR{-} &\DynN{}}
&\DynTab{\DynOD{$1$}{$1$}\DynDR{=}  &\DynOD{$4$}{$2$}\DynDDR & &\DynOD{$4$}{$2$}}
\END
%
$\mfrak{so}^*(2\rk)$
&\rule{0pt}{3em}\DynTab{                    &\DynOR{$1$}\DynD{-}\DynSym{ld}{0.8}\\
           \DynO{$1$}\DynR{-}  &\DynN{}\DynR{-} &\DynO{$2$}\DynDDR & &\DynN{}\DynR{-}
           &\DynO{$\tfrac{l-1}{2}$}\DynR{-} &\DynN{}}
&\raisebox{-2em}{\strut}\DynTab{\DynOOD{\raisebox{0pt}[\height][0pt]{\raisebox{-0.57\height}{\shortstack{$4$\\$1$}}}}{$2$}
 &\DynOD{$4$}{$2$}\DynDL{=}\DynDDR & &\DynOD{$4$}{$2$}}
\END
%
\Hline 
\ROWSP[3.3]
$\mfrak{e}_{6(-14)}$
&\rule[-1em]{0pt}{4em}\DynTab{                    &                   &\DynOR{$1$}\DynD{-}\\
                        \DynOA{$2$}\DynR{-}\DynSym{rrrr}{0.7}&\DynN{}\DynR{-} 
                        &\DynN{}\DynR{-} &\DynN{}\DynR{-} &\DynOA{$2$}}
&\raisebox{-2em}{\strut}\DynTab{           \DynOD{$6$}{$2$}\DynDR{=} &\DynOOD{\raisebox{0pt}[\height][0pt]{\raisebox{-0.57\height}{\shortstack{$8$\\$1$}}}}{$2$}}
\END
%
\Hline
$\mfrak{e}_{7(-25)}$
&\rule{0pt}{3em}\DynTab{                    &                      &\DynN{}\DynD{-}\\
           \DynO{$1$}\DynR{-}  &\DynN{}\DynR{-} &\DynN{}\DynR{-} &\DynN{}\DynR{-} 
                                                                           &\DynO{$2$}\DynR{-} &\DynO{$3$}}
&\DynTab{           \DynOD{$8$}{$2$}\DynR{-} &\DynOD{$8$}{$2$} &\DynOD{$1$}{$1$}\DynDL{=}}
\END
%
\hline \mycaption{SatakeDiag}{All Satake diagrams and corresponding
  Dynkin diagrams of the restricted root systems of simple Lie
  algebras whose maximal compact Lie subalgebra has nontrivial center.
  The Coxeter labels of restricted roots are indicated above the
  node.}
\end{mytable}

\clearpage
\begin{mytable}{2}{&}{|c|l|}{Group     & $\I\energy$}
$\LA{SU(m,m)}$ & $\Wf{\sr{m}}+\Wf{\sr{m-1}+\sr{m}+\sr{m+1}}+\cdots+\Wf{\sr{1}+\cdots+\sr{l}}$
\END
\HLINE $\LA{SO(2,\SOdim)}$ $B_l$ & $\Wf{\sr{1}}+\Wf{\rtmx}$
\END
\HLINE
$\LA{Sp(l,\R)}$ & $\Wf{\sr{1}}-\Wf{\sr{1}+2\sr{2}}+\Wf{\sr{1}+2\sr{2}+2\sr{3}}+\cdots\pm\Wf{\rtmx}$ 
\END
\HLINE
$\LA{SO(2,\SOdim)}$ $D_l$ & $\Wf{\sr{l}}+\Wf{\rtmx}$
\END
\HLINE
$\LA{SO^*(4n)}$ & $\Wf{\sr{1}}+\Wf{\sr{1}+\sr{2}+2\sr{3}+\sr{4}}+\Wf{\sr{1}+\sr{2}+2\sr{3}+2\sr{4}+2\sr{5}+\sr{6}}+\Wf{\rtmx}$
\END
\HLINE
$\LA{E_{7(-25)}}$ & $\Wf{\sr{7}}+\Wf{\sr{1}+\sr{3}+2\sr{4}+2\sr{5}+2\sr{6}+\sr{7}}+\Wf{\rtmx}$
\END
\hline

\mycaption{HICSA}{$\I\energy$ in the Cartan-Weyl basis with respect to
  the maximally non-compact Cartan subalgebra, where
  $\Wf{\rta}=\We{\rta}+\We{-\rta}$.}
\end{mytable}

\begin{proof}[ of \rst{listOfCausals}]
  \Rst{theo:1} and \rst{threegrading} requires that we should look for
  a restricted root $\srrext$ with Coxeter label 1.  Inspection shows
  that there is at most one such.  The Lie algebras of $SU(n,m)$ with
  $n<m$ and of $E_{6(-14)}$ have no restricted root with Coxeter label
  1. Therefore no space-time manifolds $\Spacetime $ exist for them
  which satisfy our assumptions.
  
  For the remaining Lie algebras, the parabolic subalgebra is uniquely
  determined and has as $\srsubset$ the set of all simple restricted
  roots other than $\srrext$. The generalized Lorentz subalgebra
  $\lorentz = \IctzI$ is generated by the simple roots which do not
  restrict to $\srrext$. The resulting $\lorentz$ are listed in
  \rfig{ParblStruct} and it can be seen that $\lorentz$ is simple in
  all cases except that $\lorentz$ is trivial in the well known case 
  $\mfrak{sl}(2,\R)\simeq\mfrak{su}(1,1)$. 

  Condition (ii) of \rst{requirements} has been checked with the help
  of the computer algebra package LambdaLie
  \cite{deRiese:LambdaLie-04} by computing $\energy$ explicitly in the
  Cartan Weyl basis based on the maximally non-compact Cartan
  subalgebra $\ICSAp\VSsum\ICSAk$. Details can be found in
  \cite{deRiese:PhD-04}. The resulting vectors $\I\energy$ can be
  found in \rfig{HICSA}. In all cases $\I\energy$ is a linear
  combination containing $\Wf{\rtmx}$ which is never in $\lorentz$.
  Therefore (ii) is satisfied for all cases.
\end{proof}

For general $\srsubset$, the structure of $\IctzI$ can also be
obtained from the Satake diagram and $\Iroots$: The subdiagram of the
Satake diagram which consists of all nodes of type 1 and the nodes of
type 2 and 3 which correspond to an element of $\Iroots$ describes the
semi-simple part of $\IctzI$.  The number of pairs of type 2 which do
not correspond to any element of $\Iroots$ is equal to the dimension
of the (possibly empty) compact Abelian ideal of $\IctzI$ -- it is
part of $\ICSAk$.

The dimension of $\Spacetime$ equals the dimension of 
$\trl$. Because of the generalized Bruhat decomposition 
\begin{equation}
  \elabel{dimInilpE}
\dim \Spacetime =
  \dim\InilpE{\pm}=\tfrac{1}{2}(\dim\rLA-\dim\IctzI-\dim\ICSApE)
\end{equation}
The results are collected in \rfig{ParblStruct}, together with the
split rank of $\rLA$, i.e. the maximal number of noncompact generators
in a Cartan subalgebra.  The split rank of the generalized Lorentz
algebra is one less, because $\ICSApE$ is 1-dimensional.

Note that the generalized Lorentz group $E_{6(-26)}$ of $E_{7(-25)}$
has split rank 2, i.e. two commuting boosts, whereas the Lorentz
groups proper has only one. This suggests a more restrictive symmetry.


\section{The Generator $\energy$ and Time Reflections}
\label{sec:HandTime}

Let us remember from the introduction that the existence of positive
energy representations of a simple Lie algebra $\rLA$ requires that
its maximal compact subalgebra $\Ck$ has a 1-dimensional center
$\Ckc$, hence $\Ck=\Cks\oplus\Ckc$, where $\Cks$ is the semi-simple
part of $\Ck$. It follows that $\rLA$ has a compact Cartan subalgebra
$\CSA$, and that the simple roots $\{\sr{1},...,\sr{\rk}\}$ with respect
to this Cartan subalgebra are the simple roots of $\Ck$ plus one simple root
$\sr{c}$ with Coxeter label 1. Read the above Satake diagrams as
Dynkin diagrams of the complex Lie algebra $\cpl{\rLA}$, with nodes
corresponding to the simple roots $\{\sr{1},...,\sr{l}\}$.  The node
corresponding to $\sr{c}$ is called the marked node.  We order the
nodes $1,\dots,l$ from top to bottom and from left to right. 
 The \rfig{ParblStruct} gives the number of the
marked node under the heading ``Node''. The Dynkin diagram of $\Cks$
is obtained by removing the marked node.

We adopt the following standard notation.  Given a linear combination
$\Lambda = \sum_i m^i\rta_i$ of simple roots, let $\Wh{\Lambda}$ be
the corresponding element of $\CSA$ which obeys
$\Lambda(\Wh{})=\Killing{\Wh{\Lambda}}{\Wh{}}$ for all $\Wh{}\in\CSA$,
where $\Killing{\cdot}{\cdot}$ is the Killing form. Its commutator
with ladder operators $\We{\sr{j}}$ of simple roots is
$\comm{\Wh{\Lambda}}{\We{\sr{j}}}=\RProd{\Lambda}{\sr{j}}\We{\sr{j}}$.
The generator $iH_0$ of the center $\Ckc$ must commute with the ladder 
operators of simple roots of $\Cks$, therefore, up to normalization
\begin{equation}
\elabel{H0}
   \energy = \I\Wh{\fw{c}} 
\end{equation}
where $\fw{c}$ is the fundamental weight which is orthogonal to all 
simple roots of $\rLA$ except $\sr{c}$. 

The fundamental weights are well known. This yields the coefficients
in the expansion of $\I\energy$ in the expansion in $\Wh{\sr{j}}$ as
given in the table below.
\begin{mytable}{3}{&&}{|c|c|c|}%
   {Name             & Index       & $\gamma \energy$                                }
    $\LA{SU(m,n)}$        & $1-(m-n)^2$ & $((1)n,\dots,(m-1)n,mn,m(n-1),\dots,m(1))$      \END
    $\LA{SO(2,d)}$, $B_l$ & $-2l^2+7l-4$& $(1,\dots,1)$                                   \END
    $\LA{Sp(l,\R)}$       & $l$         & $(\tfrac{1}{2}l,l-1,\dots,2,1)$                 \END
    $\LA{SO(2,d)}$, $D_l$ & $-2l^2+9l-8$& $(\tfrac{1}{2},\tfrac{1}{2},1,\dots,1)$         \END
    $\LA{SO^*(2l)}$       & $-l$        & $(\tfrac{1}{2}l,\tfrac{1}{2}l-1,l-2,\dots,1)$   \END
    $\LA{E_{6(-14)}}$     & -14         & $(3,2,4,6,5,4)$                                 \END
    $\LA{E_{7(-25)}}$     & -25         & $(3,2,4,6,5,4,3)$                               \END     
    \hline
    \mycaption{H}{Killing index and Time translation generator for the
      simple Lie groups with pos.  energ. representations in the basis
      corresponding to a maximally compact Cartan subalgebra. $\gamma$
      is some imaginary number.}
\end{mytable}

\subsection{Time reflection}

Let us assume a time reversal $\timerev$ exists. It acts on
$\Spacetime=\R\times \Cauchy$ in such a way that this induces an
automorphism of $\rLG$ of the form $\Homact(g)\mapsto
\timerev\Homact(g)\timerev^{-1}$ and an automorphism $\Ad(\timerev)$
of $\rLA$ with the following properties: Since $\timerev$ reflects
time, $\Ad(\timerev)\energy = -\energy$. Since $\timerev$ should act
trivially on space $\Cauchy$, it has to be $\Ad(\timerev)X = X $ for
all $X\in \Cks$.

Recall from \req{H0} that $\energy=\I\Wh{\fw{i}}$, where $i$ is the number of the
marked node. This clearly is a linear combination of the $\Wh{\sr{j}}$
with pure imaginary coefficients.

\begin{lemma}
Write the generator of time translations in the form
\begin{equation}
\elabel{Hroot}
 \energy = r \Bigl( \Wh{\sr{i}} + \tfrac{1}{2}\sum_{j\neq i}\mu_j \Wh{\sr{j}}\Bigr) \ .
\end{equation}
with pure imaginary $r$.

An automorphism $\Ad(\timerev)$ of $\rLA$ such that
$\Ad(\timerev)\energy=-\energy$ and $\Ad(\timerev)\Wh{\sr{j}} =
\Wh{\sr{j}}$ for $j\neq i$ can only exist if $\mu_j$ are integers.
\end{lemma}

\begin{proof}
  The automorphism $\Ad(\timerev)$ of $\rLA$ maps the Cartan
  subalgebra $\CSA$ into itself and must therefore define an
  automorphism of the root system. We will also call it $\timerev$. In
  particular, $\timerev(\sr{j})$ must be a root, $T(\fw{i})=-\fw{i}$,
  because of \req{H0}, and $\timerev(\sr{j})= \sr{j}$
  for $j\neq i$.
 
  From \req{Hroot} it follows that
  \begin{equation}\elabel{}
    \timerev(\sr{i}) = \bigl( - \sr{i} - \sum_{j\neq i} \mu_j \sr{j}\Bigr)\,.
  \end{equation} 
  Roots are sums of integer multiples of simple roots, therefore
  $\timerev(\sr{i})$ is not a root if any of the $\mu_j$ are
  non-integer.
\end{proof}  

Using this result and \rfig{H}, necessary conditions for the 
existence of a time reflection automorphism are obtained. It is well known and 
compatible with these necessary conditions that $\LA{SO(2,d)}$ admits a time
 reflection automorphism. Apart from these, only the two Lie algebras 
$\LA{SU(1,2)}$ and $\LA{Sp(4,\R)}$ satisfy the necessary conditions. 
Using the computer algebra package LambdaLie it was found that 
$\LA{SU(1,2)}$ admits a time reflection automorphism, but
$\LA{Sp(4,\R)}$ does not. Details are as follows.

\begin{proof}[ of \rst{theo:timereversal}]
Examine the necessary conditions.
 For $\LA{SU(m,n)}$ the following have to be
integer: $\frac{2p}{m}$ for $0<p<m$ and $\frac{2q}{n}$ for $0<q<n$.
Therefore $m,n\le2$. The only case not covered by an isomorphic
conformal group is $\LA{SU(2,1)}$. For $\LA{Sp(l,\R)}$,
 $\frac{4p}{l}$ has to be
integer for $0<p<l$, which is true only for $l=1,2,4$ where only the
last case is not isomorphic to a conformal group. For $\LA{SO^*(2l)}$
$\frac{2l-4}{l}$ and $\frac{4p}{l}$ for $0<p<l-1$ have to be integer.
This is true for $l=2,4$, but $\LA{SO^*(4)}$ is not simple and $\LA{SO^*(8)}$
again isomorphic to a conformal group. For $\LA{E_{6(-14)}}$ and
$\LA{E_{7(-25)}}$ we have $\mu=(\frac{3}{2},1,2,3,\frac{5}{2},1)$ and
$\mu=(2,\frac{2}{3},\frac{4}{3},2,\frac{5}{3},\frac{4}{3},1)$
respectively. Thus, both do not allow a time reversal.

This leaves the two cases $\LA{SU(1,2)}$ and $\LA{Sp(4,\R)}$ which are
settled by computer algebra as explained in \cite{deRiese:PhD-04}.
\end{proof}

\section{Outlook}

In conformal field theory in 4 dimensions, further results were
obtained which it would be interesting to generalize.

\begin{enumerate}
\item The irreducible positive energy representations of $\rLG$ can be
  constructed as induced representations on $\Spacetime$
  \cite{Mack:COMMP-55-1}. They are induced by finite dimensional
  representations of the parabolic stability group
  $\Homstab=\Lorentz\Dil\Spc$ which are trivial on $\Spc$, and are
  labelled $(l,\delta)$ where $l$ specifies a finite dimensional
  representation of $\Lorentz$, and the dimension $\delta$ specifies
  the representation of $\Dil$. Since in an irreducible representation
  the center is represented by a multiple of the identity, the
  functions $\Psi_\alpha(x)$ in the representation space actually live
  on $\quotient{\Spacetime}{\Z}$, where $\Z $ is the $\infty$ factor
  of the center.  The scalar product is furnished by the Kunze Stein
  formula \cite{KunzeStein:AJM-67} and the only nontrivial task is to
  determine when it is positive, and to treat some degenerate cases.
  
  It is essential for this construction that $H$ is (the identity
  component of) a parabolic subgroup.
\item One introduces field operators $\phi_\alpha(x), x\in \Spacetime$
  with corresponding transformation law labelled by $(l,\delta)$ and
  such that $\phi_\alpha(x)|\Omega\rangle $, $\Omega =$ vacuum span
  irreducible positive energy representation spaces isomorphic to the
  above.  The two point function $\langle \Omega | \phi_\beta^\ast
  (y)\phi_\alpha(x)|\Omega \rangle$ equals the above scalar product.
\item One shows that the 3-point functions $\langle \Omega |
  \phi_\alpha(x) \Phi_\beta (y) \phi_\gamma (z)|\Omega \rangle $ are
  determined by symmetry up to some arbitrary (coupling) constants
  \cite{Polyakov:old}.
\item Suitably summed up operator product expansions
  $\phi_\alpha(x)\phi_\beta(y)|\Omega\rangle =\dots $ converge on the
  vacuum $\Omega$ if they are valid as short distance expansions, because they
  amount to partial wave expansions on the conformal group~\cite{Mack:COMMP-53-155}.
\item Given that this is so, one can construct $n$-point functions
  $\langle \Omega |\phi_{\alpha_1}(x_1)\dots$ $\dots\phi_{\alpha_n}(x_n)|\Omega
  \rangle$ from the two and three point functions. The will satisfy
  all Wightman axioms \cite{StreaWight:PCT-64} except local
  commutativity
  \begin{equation}
  \elabel{commutativity}
    \comm{\phi_\alpha(x)}{\phi_\beta (y)}_- = 0
  \end{equation} 
  of observable fields such as the stress energy tensor, for
  relatively spacelike $x,y$.  They depend on the afore mentioned a
  priori arbitrary coupling constants.
\item To construct a local theory, the remaining nontrivial task is
  then to satisfy the commutation relations \req{commutativity}. If
  it should turn out that for some groups $\rLG$ also the four point
  functions are determined by symmetry (up to some normalization
  constants), as is the case in 2-dimensional conformal field theory,
  this task would be very much simplified.
\item The relation between conformal field theory in the universal
  covering of compactified Minkowski space and in Euklidean space was
  established by L\"uscher and Mack
  \cite{LueMack:COMMP-41-203,Luscher:DESY-75-51}.  It involves
  analytic continuation through a holomorphic semi-group which acts
  contractively on the Hilbert space of physical states.  The method
  involves Euklidean time reversal as an involutive automorphism.
  Similar methods are employed in the Gel'fand-Gindikin program
  \cite{GelfandGindikin:CM-78}, which aims at the realization of
  families of similar unitary representations of Lie groups in a
  unified geometric way, and was applied to causal symmetric spaces
  \cite{Olafsson:ACRTHA-00,BieliavskyPevzner:SSSR2CSS-01}.
  
  We expect that in general the relation of $\rLG$ with a Euklidean
  symmetry will be more subtle than in the conformal case.
\item It 
is
interesting to study supersymmetric versions of the
  pairs $(G,\Spacetime)$ as generalizations of super conformal field
  theory. 
Jordan super algebras and their associated super conformal groups 
were classified by G\"unaydin \cite{GSUSY1,GSUSY2}. He also presented a general
 way to construct unitary positive energy representations, for both the Lie 
algebras and super Lie algebras, 
by using  multiplets of step operators for harmonic oscillators
 \cite{GOszRep1,GOszRep2,MackTodorov}. 
   
\end{enumerate}
The  identification of lowest weigt representations with induced representations for all our groups $G$ was already accomplished  by 
G\"unaydin, using the isomorphism mentioned in \rsec{obs}
 He used this to illuminate  AdS/CFT correspondences from a representation
 theoretic point of view.  

A group theoretical study of dimensional reduction would also be of
interest.  In the present context, space is always compact (a sphere
in the conformal case). In Minkowski space, decompactification is
associated with the breaking of conformal symmetry. In particular,
mass generation effectively removes the point at $\infty$.

The exceptional group $E_{7(-25)}$ is particularly interesting. It
contains a maximal subgroup $G^4\times \tilde{U}$ where $G^4=SO(2,4)$
may be interpreted as conformal space-time symmetry in 4 dimensions,
and $\tilde{U}=SU(4)\times SL(2,\R)$ which one is tempted to interpret
as an internal symmetry. The strange feature is the appearance of a
noncompact internal symmetry group $SL(2,\R)$, reminiscent of hidden
symmetries in supergravity
\cite{Koepsell:MSES-01,GKN:CQRELG-01,GKN:MUR-01}.

It would appear at first sight that we will not quite get the $
SU(3)\times U(1)\times SU(2)$-symmetry group of the standard model,
because there is a factor $SL(2,\R)$ in place of $SU(2)$

However, the internal symmetry of the dimensionally reduced theory is
more subtle, because

i) The internal symmetry which acts nontrivially may be smaller than 
$\tilde{U}$.

ii) There is no difference between compact and non-compact real forms
for internal symmetries which act on finite-dimensional vector spaces
(e.g. act on indices of fields), because of a corollary of Weyl's
unitary trick which asserts not only equivalence of representations
but also of invariants (see Appendix A).  Every semi-simple complex
group $U_\C$ possesses a compact real form.

\com{********* Cite Koepsell somewhere, check Name spelling below ************}

It is also of interest to consider space-time manifolds $\Spacetime =
\quotient{\rLG}{\Homstab}$ with semi-simple or reductive stability
group $\Homstab$. $\rLG$-invariant causal structures on symmetric
spaces $\quotient{\rLG}{\Homstab}$ were investigated in the
mathematical literature
\cite{HilOlaf:CSS-97,FarautOlafsson:CSSSGHA-95}.  This includes anti
de Sitter space.  We did not deal with this case in the body of the
paper, but we list some examples with semi-simple $\rLG$ which
obviously satisfy the requirement of causality.

Let $\rLG_1$ be a simple connected simply connected Lie group which
possesses positive energy representations, i.e. the universal covering
of one of the groups in \req{groupsWithPosEnergyReps}. 
\begin{del}L\"uscher has shown \cite{Luscher:DESY-75-51} \end{del}%
\begin{corr}From arguments of L\"uscher \cite{Luscher:DESY-75-51} it follows 
\end{corr}%
that $\rLG_1$ possesses an $ \Ad(\rLG_1)$-invariant causal structure.
Set $\Spacetime=\rLG_1$, acted upon by elements
$(g_L,g_R)\in\rLG=\rLG_1\times\rLG_1$ according to $ m\mapsto
g_Lmg_R^{-1}$.  Then $\Homstab$ is the diagonal subgroup of $G_1\times G_1$
and its elements $(g,g)$ act on $\Spacetime$ according to $m\mapsto
\Ad(g)m$.  It follows from elementary results in the body of this
paper that a $\Ad(\rLG_1)$ invariant causal structure on $\Spacetime $ is
also $\rLG$-invariant.

The universal covering of the central extension of the diffeomorphism
group $\rLG_1$ of the circle admits positive energy representations.
The Lie algebra of $\rLG=\rLG_1\times\rLG_1$ consists of two copies of
the Virasoro algebra.  Assuming that L\"uschers arguments extend to
this $\infty$-dimensional example, we would conclude that the
$\infty$-dimensional space-time manifold $\Spacetime=\rLG_1$ admits a
$\rLG$-invariant causal structure. This suggests that quantum field
theories on $\infty$-dimensional space-time manifolds may exist.

\subsection*{Acknowledgement}

We are grateful to Kilian Koepsell for explaining his work, and to
Claudia Lehmann and Thomas Fischbacher for discussions. M. de R.
thanks Hermann Nicolai for an invitation and discussions and the
German National Academic Foundation for financial support.

\section*{Appendix A: Weyl's unitary trick} 

\begin{proposition}
  Let $G^\prime $ and $\rLG$ be two real forms of a complex Lie group
  $G_\C$, with Lie algebras $\rLA^\prime =\rLA_0 \oplus \rLA_1 \subset
  \rLA_\C$ and $\rLA =\rLA_0 \oplus i\rLA_1 \subset \rLA_\C$.  Let
  $V^0,V^1,\dots ,V^n$ be finite dimensional complex representation
  spaces carrying representations $\tau^1,...,\tau^n$ of $G^\prime $.
  Then
  \begin{itemize}
  \item [i.] The representation operators $\tau^i$ qua functions of
    $g\in G^\prime$ extend to holomorphic functions on $G_\C$, making
    $V^i$ into representation spaces for $G_\C$, hence of $\rLG$.
  \item [ii.] Every intertwiner 
    \begin{equation} 
    \elabel{}
      C: V^1\otimes ...\otimes
      V^n\mapsto V^0 
    \end{equation} 
    between representations of $G^\prime$ extends to an intertwiner of
    representations of $G_\C$, hence of $\rLG$.
  \end{itemize}
\end{proposition}

Invariants are intertwiners to the trivial representation.

\begin{proof}
  part i. is known as Weyl's unitary trick.
  
  The intertwining property reads
  \begin{equation}
    \elabel{}
    C\bigl( \tau^1(g)\otimes ... \otimes \tau^n(g)\bigr)
    = \tau^0(g)C
  \end{equation}
  for $g\in G^\prime$. By i., both sides extend to holomorphic
  functions on $G_\C$ which agree on the real neighborhood $G^\prime$.
  They are therefore equal.
\end{proof}

\bibliography{mrabbrev,bibSPIRES}


\end{document}